\documentclass[sigplan,screen]{acmart}
\usepackage{microtype}
\usepackage{graphicx}
\usepackage{adjustbox}
\usepackage{amsmath}
\usepackage[utf8]{inputenc}
\usepackage{ccicons}
\usepackage{verbatim}
\usepackage{amscd}
\usepackage{xcolor}
\usepackage{bbold}
\usepackage{url}
\usepackage{upgreek}
\usepackage{stmaryrd}
\usepackage{lipsum}
\usepackage{tikz-cd}
\usetikzlibrary{cd}
\usetikzlibrary{calc}
\usetikzlibrary{arrows}
\usepackage{bussproofs}
\EnableBpAbbreviations
\DeclareMathAlphabet{\mathpzc}{OT1}{pzc}{m}{it}

\newcommand{\List}{\mathsf{List}\,}

\newcommand{\Set}{\mathsf{Set}}

\newcommand{\ol}[1]{\overline{#1}}

\copyrightyear{2022}
\acmYear{2022}
\setcopyright{acmlicensed}\acmConference[CPP '22]{Proceedings of the 11th
ACM SIGPLAN International Conference on Certified Programs and
Proofs}{January 17--18, 2022}{Philadelphia, PA, USA}
\acmBooktitle{Proceedings of the 11th ACM SIGPLAN International Conference
on Certified Programs and Proofs (CPP '22), January 17--18, 2022,
Philadelphia, PA, USA}
\acmPrice{15.00}
\acmDOI{10.1145/3497775.3503680}
\acmISBN{978-1-4503-9182-5/22/01}

\begin{document}
\title[(Deep) Induction Rules for GADTs]{(Deep) Induction Rules for GADTs}
  \author{Patricia Johann}
  \email{johannp@appstate.edu}
\affiliation{\institution{Appalachian State University}\country{USA}}
  \author{Enrico Ghiorzi}
  \authornote{New address: Istituto Italiano di Tecnologia, Italy,
      enrico.ghiorzi@iit.it}
  \email{ghiorzie@appstate.edu}
\affiliation{\institution{Appalachian State University}\country{USA}}

\begin{abstract} 
  Deep data types are those that are constructed from other data types,
  including, possibly, themselves.  In this case, they are said to be
  truly nested.  Deep induction is an extension of structural
  induction that traverses {\em all} of the structure in a deep data
  type, propagating predicates on its primitive data throughout the
  entire structure.  Deep induction can be used to prove properties of
  nested types, including truly nested types, that cannot be proved
  via structural induction.  In this paper we show how to extend deep
  induction to GADTs that are not truly nested GADTs.  This opens the
  way to incorporating automatic generation of (deep) induction rules
  for them into proof assistants.  We also show that the techniques
  developed in this paper do not suffice for extending deep induction
  to truly nested GADTs, so more sophisticated techniques are needed
  to derive deep induction rules for them.
\end{abstract}




\maketitle

\section{Introduction}\label{sec:intro}

Induction is one of the most important techniques available for
working with advanced data types, so it is both inevitable and
unsurprising that it plays an essential role in modern proof
assistants. In the proof assistant Coq~\cite{coq20}, for example,
functions and predicates over advanced types are defined inductively,
and almost all non-trivial proofs of their properties are either
proved by induction outright or rely on lemmas that are.  Every time a
new inductive type is declared in Coq an induction rule is
automatically generated for it.\looseness=-1

The inductive data types handled by Coq include (possibly mutually
inductive) polynomial algebraic data types (ADTs), and the induction
rules Coq generates for them are the expected ones for standard
structural induction. However, as discussed in~\cite{jp20}, it has
long been understood that these rules are too weak to be genuinely
useful for deep ADTs.\footnote{An ADT/nested type/GADT is {\em deep}
  if it is (possibly mutually inductively) defined in terms of other
  ADTs/nested types/GADTs (including, possibly, itself).}  The
following data type of rose trees, here coded in Agda and defined in
terms of the standard type $\mathsf{List}$ of lists (see
Section~\ref{sec:ADTs-and-nesteds}), is a deep ADT:
\[\begin{array}{l}
\mathsf{data\, Rose\, : Set \to Set\,where}\\
\mathsf{\;\;\;empty :\,\forall \{A : Set\} \to Rose\,A}\\
\mathsf{\;\;\;node\,\,\,\, :\, \forall \{A : Set\} \to A \to List\,(Rose\,A) \to Rose\,A} 
\end{array}\]
The induction rule Coq automatically generates for (the analogous Coq
definition of) rose trees is
\[\begin{array}{l}
\mathsf{\forall\, (A : Set)\,(P : Rose\,A \to Set) \to}\\
\mathsf{\hspace*{0.15in}P\,empty \to}\\
\mathsf{\hspace*{0.15in}
 (\forall\, (a : A)\,(ts :
  List\,(Rose\,A)) \to P\,(node\,a\,ts)) \to}\\ 
\hspace*{0.15in}  \mathsf{\forall \,(x : Rose\,A) \to P\,x}
\end{array}\]
\noindent
Unfortunately, this is neither the induction rule we intuitively
expect, nor is it expressive enough to prove even basic properties of
rose trees that ought to be amenable to inductive proof. What is
needed here is an enhanced notion of induction that, when specialized
to rose trees, will propagate the predicate $\mathsf{P}$ through the
outer list structure and to the rose trees sitting inside
$\mathsf{node}$'s list argument. More generally, this enhanced notion
of induction should traverse {\em all} of the layers present in a data
structure, propagating suitable predicates to {\em all} of the data it
contains.  With data types becoming ever more advanced, and with
deeply structured types becoming increasingly ubiquitous in
formalizations, such an enhanced notion of induction is essential if
proof assistants are to be
%it is critically important that proof assistants
%%{\color{red} (e.g., Coq, Agda, Lean)}
%are
able to automatically generate genuinely useful induction rules for
data types that go
%well
beyond traditional ADTs. These
%Such data types
include not just deep ADTs, but also (truly\footnote{A truly nested
  type is a nested type that is defined over itself. The data type
  $\mathsf{Bush}$ in Section~\ref{sec:ADTs-and-nesteds} provides a
  concrete example.}) nested types~\cite{bm98},
%, such as $\mathsf{Rose}$ above, as well as truly nested types
generalized algebraic data types
(GADTs)~\cite{ch03,pvww06,sp04,xcc03}, more richly indexed
families~\cite{dyb94}, and deep variants of all of these. A summary of
the various classes of data types considered in this paper is given in
Table~\ref{fig:taxonomy}.\looseness=-1

\begin{table}[h!]
\begin{tabular}{|l|l|l|}
\hline
Data types & Discussed & Examples \\
          & in Sections &          \\
\hline\hline
ADTs & \ref{sec:intro} &
$\mathsf{List}$, $\mathsf{Rose}$ \\ 
Nested types & \ref{sec:ADTs-and-nesteds} & $\mathsf{PTree}$ \\  
Truly nested types & \ref{sec:ADTs-and-nesteds} & $\mathsf{Bush}$\\
GADTs & \ref{sec:GADTs} & $\mathsf{Eq}$, $\mathsf{Seq}$\\
Truly nested GADTs & \ref{sec:GADTs} and~\ref{sec:GADT-nested}&
$\mathsf{G}$ in \eqref{gadt-nested}\\ 
\hline
\end{tabular}
\caption{Data types in this paper}\label{fig:taxonomy}
\end{table}

{\em Deep induction}~\cite{jp20} is a generalization of structural
induction that fits this bill exactly. Whereas structural induction
rules induct over only the top-level structure of data, leaving any
data internal to the top-level structure untouched, deep induction
rules induct over {\em all} of the structured data present. The key
idea is to parameterize induction rules not just over a predicate over
the top-level data type being considered, but also over additional
custom predicates on the types of primitive data they contain. These
custom predicates are then lifted to predicates on any internal
structures containing these data, and the resulting predicates on
these internal structures are lifted to predicates on any internal
structures containing structures at the previous level, and so on,
until the internal structures at all levels of the data type
definition, including the top level, have been so
processed. Satisfaction of a predicate by the data at one level of a
structure is then conditioned upon satisfaction of the appropriate
predicates by {\em all} of the data at the preceding level.

Deep induction was shown in~\cite{jp20} to deliver induction rules
appropriate to nested types, including ADTs.  For the (deep) ADT of
rose trees, for example, it gives the following genuinely useful
induction rule:
\begin{equation}\label{eq:rose}
\begin{array}{l}
\mathsf{\forall \,(A : Set)\,(P : Rose\,A \to Set)\,(Q : A \to Set)
  \to}\\
\mathsf{\hspace*{0.15in}P\,empty \to}\\ 
\mathsf{\hspace*{0.15in}(\forall \,(a : A)\, (ts :
  List\,(Rose\,A))\to Q\,a \to}\\
\mathsf{\hspace*{0.3in}List^\land\,(Rose\,A)\, P\,ts \to P\,(node\,a\,ts)) \to}\\
\mathsf{\hspace*{0.15in}\forall \,(x :
  Rose\,A) \to Rose^\land\,A\,Q\, x \to P\,x} 
\end{array}
\end{equation}
Here, $\mathsf{List^\land}$ (resp., $\mathsf{Rose^\land}$) lifts its
predicate argument $\mathsf{P}$ (resp., $\mathsf{Q}$) on data of type
$\mathsf{Rose\,A}$ (resp., $\mathsf{A}$) to a predicate on data of
type $\mathsf{List\,(Rose\,A)}$ (resp., $\mathsf{Rose\,A}$) asserting
that $\mathsf{P}$ (resp., $\mathsf{Q}$) holds for every element of its
list (resp., rose tree) argument.\footnote{Predicate liftings such as
  $\mathsf{List^\land}$ and $\mathsf{Rose^\land}$ can either be
  supplied as primitives or generated automatically from their
  associated data type definitions as described in
  Section~\ref{sec:ADTs-and-nesteds}. The predicate lifting for a
  container type like $\mathsf{List\,A}$ or $\mathsf{Rose\,A}$ simply
  traverses containers of that type and applies its predicate argument
  pointwise to the constituent data of type $\mathsf{A}$.  The ability
  to define predicate liftings for more general data types will be
  critical to deriving their deep induction rules in
  Section~\ref{sec:framework}.} Deep induction was also shown
in~\cite{jp20} to
% be the missing piece making it possible to
%be the theretofore missing component needed to
% be the essential ingredient making it possible to
deliver the first-ever induction rules --- structural or otherwise ---
for the $\mathsf{Bush}$ data type~\cite{bm98} and other truly nested
types. Deep induction for ADTs and (truly) nested types is reviewed in
Section~\ref{sec:ADTs-and-nesteds}.

This paper shows how to extend deep induction to proper GADTs, i.e.,
GADTs that are not nested types (and thus are not ADTs).  Typical
applications of such GADTs include generic programming, modeling
programming languages via higher-order abstract syntax, maintaining
invariants in data structures, and expressing constraints in embedded
domain-specific languages. They have also been used to implement
tagless interpreters~\cite{pl04,pr06,pvww06} by trading the definition
of a universal value domain for a direct specification of the property
of being a value. Other applications are described in,
e.g.,~\cite{min15,rou06}. A constructor for a GADT $\mathsf{G}$ may,
like a constructor for a nested type, take as arguments data whose
types involve instances of $\mathsf{G}$ other than the one being
defined. These can even include instances involving $\mathsf{G}$
itself. But if $\mathsf{G}$ is a proper GADT, then at least one of its
constructors will also have a structured instance of $\mathsf{G}$ ---
albeit one not involving $\mathsf{G}$ itself --- as its codomain.
%But while the types of the arguments to a GADT's constructor can
%include instances of $\mathsf{G}$ that themselves involve
%$\mathsf{G}$, its return type cannot.
For example, the constructor $\mathsf{pair}$ for the GADT 

\vspace*{-0.1in}

\begin{equation}\label{eq:seq}
\begin{array}{l}
\mathsf{data\, Seq\,: Set \to Set\,where}\\
\mathsf{\;\;const :\, \forall \{A : Set\} \to A \to Seq\,A}\\
\mathsf{\;\;pair\;\;\, :\,\forall \{A\,B : Set\} \to Seq \,A \to Seq\,B \to
  Seq\,(A \times B)}
\end{array}
\end{equation}
\noindent
of sequences\footnote{The type of $\mathsf{Seq}$ is actually
  $\mathsf{Set \to \Set_1}$, but to aid readability we elide the
  explicit tracking of Agda universe levels in this paper.} only
constructs sequences of data whose types are pair-structured, rather
than sequences of arbitrary type, as does $\mathsf{const}$. If one or
more of the data constructors for a GADT $\mathsf{G}$ return
structured instances of $\mathsf{G}$, then the GADT will have two
distinct, but equally natural, semantics: a functorial semantics
interpreting it as a left Kan extension~\cite{mac71}, and a parametric
semantics interpreting it as the interpretation of its Church
encoding~\cite{atk12,vw10}. As detailed in~\cite{jg21}, a key
difference in the two semantics is that the former views GADTs as
their {\em functorial completions}~\cite{jp19}, and thus as containing
more data than just those expressible in syntax. By contrast, the
latter views them as what might be called {\em syntax-only}\/
GADTs. Fortunately, these two views of GADTs coincide for those GADTs
that are ADTs or (other, including truly) nested types.  However, both
they and their attendant properties differ greatly for proper
GADTs. In fact, the functorial and parametric semantics for proper
GADTs are sufficiently disparate that, by contrast with the semantics
customarily given for ADTs and nested
types~\cite{bfss90,gjfor15,jgj21f}, it is not at all clear how to
define a functorial parametric semantics for GADTs~\cite{jg21}.

This observation seems, at first, to be a death knell for the prospect
of extending deep induction to GADTs. Indeed, induction can be seen as
unary parametricity, so GADTs viewed as their functorial completions
do not obviously support induction rules.  This makes sense
intuitively: induction is a syntactic proof technique, so it may not
be possible to use it to prove properties of those elements of a
GADT's functorial completion that are not expressible in syntax. All
is not lost, however. As we show below, the syntax-only view of GADTs
determined by their Church encodings {\em does} support induction
rules --- including deep induction rules --- for GADTs. Indeed, this
paper gives the first-ever deep induction rules
%--- deep or otherwise ---
for proper GADTs.
%It also shows how the structural induction rule for
%a GADT can be derived as an instance of its deep induction rule.
But it actually delivers far more: it gives a {\em general framework}
for deriving deep induction rules for GADTs that can be instantiated
to particular GADTs of interest. This framework can serve as a basis
for extending modern proof assistants' automatic generation of
structural induction rules for ADTs to automatic generation of deep
induction rules for GADTs. In addition, as for ADTs and nested types,
the structural induction rule for any GADT can be recovered from its
deep induction rule by taking the custom predicates in its deep
induction rule to be constantly $\mathsf{True}$-valued (i.e.,
constantly $\mathsf{\top}$-valued) predicates.

Significantly, deep induction rules for GADTs cannot be derived by
somehow extending the approach of~\cite{jp20} to syntax-only
GADTs. Indeed, the approach taken there makes crucial use of the
functoriality of data types' interpretations from~\cite{jp19}, and
functoriality is precisely what interpreting GADTs as the
interpretations of their Church encodings fails to deliver;
see~\cite{jg08} for a discussion of why $\mathsf{Seq}$, e.g., is not
functorial. Our approach is to instead first give a predicate lifting
styled after those of~\cite{jp20}, together with a (deep) induction
rule, for the simplest --- and arguably most important --- GADT,
namely the equality GADT~\eqref{eq:equal}. We then derive the deep
induction rule for a more complex GADT $\mathsf{G}$ by {\em i}) using
the equality GADT to represent $\mathsf{G}$ as its so-called {\em
  Henry Ford encoding}~\cite{ch03,hin03,mcb99,sjsv09,sp04}, and {\em
  ii}) using the predicate liftings for the equality GADT and the
other GADTs appearing in the definition of $\mathsf{G}$ to
appropriately thread the custom predicates for the primitive types
appearing in $\mathsf{G}$ throughout $\mathsf{G}$'s structure. This
two-step process delivers deep induction rules for a very general
class of GADTs. To illustrate, we introduce a series of increasingly
complex GADTs as running examples in Section~\ref{sec:GADTs} and
derive a deep induction rule for each of them in
Section~\ref{sec:deep-ind-GADTs}. In particular, we derive the deep
induction rules for the equality data type in
Section~\ref{sec:ind-equal} and the $\mathsf{Seq}$ data type
in~\eqref{eq:seq} in Section~\ref{sec:ind-seq}. We present our general
framework for deriving (deep) induction rules for GADTs in
Section~\ref{sec:framework}, and observe that the derivations in
Section~\ref{sec:deep-ind-GADTs} are all instances of it.  In
Section~\ref{sec:GADT-nested} we show that, by contrast with truly
nested types, which do have a functorial semantics, syntax-only GADTs'
lack of functoriality means that it is not clear how to extend
induction --- deep or otherwise --- to {\em truly nested GADTs}, i.e.,
to proper GADTs whose recursive occurrences appear below
themselves.\footnote{Note carefully the distinction between a GADT
  that is not a nested type --- i.e., a proper GADT --- and a proper
  GADT that is not a truly nested GADT.  In fact, truly nested types
  are not proper GADTs and truly nested GADTs are not (truly) nested
  types. There is ample scope for confusion in light of the original,
  and now well-established, use of the term ``nested type'' to refer
  to {\em any} type that allows non-variable instances in the domains
  of its constructors, whether or not that type involves actual
%  types that do not necessarily involve actual
  nesting~\cite{bm98}.}  This does not appear to be much of a
restriction, however, since truly nested GADTs do not, to our
knowledge, appear in practice or in the
literature. Section~\ref{sec:app} comprises a case study in using deep
induction. All of the deep induction rules appearing in this paper
have been derived by instantiating our general framework. Our Agda
implementation of them is available at
\url{https://cs.appstate.edu/~johannp/CPP22Code.html}.
\looseness=-1
%at~\cite{web-page}.

\vspace*{0.05in}

{\bf Additional Related Work\/} Various techniques for deriving
induction rules for data types that go beyond ADTs have been
studied. For example, Fu and Selinger~\cite{fs18} show, via examples,
how to derive induction rules for arbitrary nested
types. Unfortunately, however, their technique is rather {\em ad hoc},
so is unclear how to generalize it to nested types other than the
specific ones studied there. Moreover,~\cite{fs18} actually derives
induction rules for data types {\em related} to the original nested
types rather than for the original nested types themselves, and it is
unclear whether or not the derived rules are sufficiently expressive
to prove all results about the original nested types that we would
expect to be provable by induction. This latter point echoes the issue
with Coq-derived induction rule for rose trees raised in
Section~\ref{sec:intro}, which has the unfortunate effect of forcing
users to manually write induction (and other) rules for such types for
use in that system. Tassi~\cite{tas19} derives induction rules for
data type definitions in Coq using unary parametricity. His technique
seems to be essentially equivalent to that of~\cite{jp19} for nested
types, although he does not permit true nesting. More recently,
Ullrich~\cite{ull20} has implemented a plugin in MetaCoq to generate
induction rules for nested types. This plugin is also based on unary
parametricity, and true nesting still is not permitted.  As far as we
know no attempts have been made to extend either implementation to
truly nested types or to proper GADTs or their deep variants.  Other
systems, including Isabelle and Lean, also derive induction rules for
data types that go beyond ADTs.  But we know of no work other than
that reported here that specifically addresses induction rules for the
(deep) GADTs considered in this paper.

\section{Deep Induction for ADTs and Nested Types}\label{sec:ADTs-and-nesteds}

A structural induction rule for a data type allows us to prove that if
a predicate holds for every element inductively produced by the data
type's constructors then it holds for every element of the data type.
In this paper, we are interested in induction rules for proof-relevant
predicates.  A proof-relevant predicate on $\mathsf{A : Set}$ is a
function $\mathsf{P\,:\,A \to Set}$ mapping each $\mathsf{a : A}$ to
the set of proofs that $\mathsf{P\,a}$ holds.  For example, the
structural induction rule for the list type
\begin{equation*}\label{eq:list}
\begin{array}{l}
\mathsf{data\ List : Set \to Set\ where}\\
\mathsf{\;\;nil\,\,\,\,\,\; :\, \forall \{A : Set\} \to List\,A}\\
\mathsf{\;\;cons\, :\, \forall \{A : Set\} \to A \to List\,A \to List\,A} 
\end{array}
\end{equation*}
is
\begin{equation*}
  \begin{array}{l}
  \mathsf{\forall (A : Set) (P : List\,A \to Set) \to}\\
  \mathsf{\hspace*{0.15in}P\,nil \to}\\
  \mathsf{\hspace*{0.15in}\big( \forall (a : A) (as: List\,A) \to P\,as
    \to P\,(cons\,a\,as)\big) \to}\\
  \mathsf{\hspace*{0.15in}\forall (as : List\,A) \to P\, as}
\end{array}
\end{equation*}
As in Coq's induction rule for rose trees, the data inside a structure
of type $\mathsf{List}$ is treated monolithically (i.e., is ignored)
by this structural induction rule.  By contrast, the deep induction
rule for lists is parameterized over a custom predicate $\mathsf{Q}$
on $\mathsf{A}$. For $\mathsf{List^\wedge}$ as described in the
introduction the deep induction rule for lists is
\[\begin{array}{l}
\mathsf{\forall (A : Set) (P : List\, A \to Set) (Q : A \to Set)
  \to}\\
\mathsf{\hspace*{0.15in}P\,nil \to}\\
\mathsf{\hspace*{0.15in}\big( \forall (a : A) (as: List\,A) \to Q\,a \to P\,as
\to P\,(cons\,a\,as)\big) \to} \\ 
\mathsf{\hspace*{0.15in}\forall (as : List\,A) \to List^{\wedge}\,A\,Q\,as
  \to P\,as } 
\end{array}\]

Structural induction can be extended to nested types, such as the
following type of perfect trees~\cite{bm98}:
\begin{equation*}\label{eq:ptree}
\begin{array}{l}
\mathsf{data\ PTree : Set \to Set\ where}\\
\mathsf{\;\;pleaf\,\,\; :\,\forall \{A : Set\} \to A \to PTree\,A}\\
\mathsf{\;\;pnode\, :\,\forall \{A : Set\} \to PTree\,(A \times A) \to PTree\,A} 
\end{array}
\end{equation*}
Perfect trees can be thought of as lists constrained to have lengths
that are powers of 2. In the above code, the constructor
$\mathsf{pnode}$ uses data of type $\mathsf{PTree\,(A \times A)}$ to
construct data of type $\mathsf{PTree\,A}$. Thus, it is clear that the
instances of $\mathsf{PTree}$ at various indices cannot be defined
independently, and that the entire inductive family of types must
therefore be defined at once. This intertwinedness of the instances of
nested types is reflected in their structural induction rules, which,
as explained in~\cite{jp20}, must necessarily involve polymorphic
predicates rather than the monomorphic predicates appearing in
structural induction rules for ADTs. The structural induction rule for
perfect trees, for example, is
\[\begin{array}{l}
\mathsf{\forall (P : \forall (A : Set) \to PTree\, A \to Set) \to}\\
\mathsf{\hspace*{0.15in}\big( \forall (A : Set) (a : A) \to P\,A\,(pleaf\, a) \big) \to}\\
\mathsf{\hspace*{0.15in}\big( \forall (A : Set) (pp : PTree\,(A \times A)) \to}\\
\mathsf{\hspace*{0.3in}P\,(A \times A)\,pp \to P\,A\,(pnode\,pp)\big) \to}\\
\mathsf{\hspace*{0.15in}\forall (A : Set) (p : PTree\,A) \to P\,A\,p} 
\end{array}\]

\begin{figure*}[t]

\begin{adjustbox}{varwidth=7.3in, max width=7.1in, margin=-0.0in 0in
      -0.2in 0in, center} 

{\small
\[\begin{array}{l}
\mathsf{\forall (P : \forall (A : Set) \to (A \to Set) \to PTree\,A
  \to Set) \to \big( \forall (A : Set) (Q : A \to Set) (a : A) \to
  Q\,a \to P\,A\,Q\,(pleaf\, a) \big) \to} \\ \quad \mathsf{ \big(
  \forall (A : Set) (Q : A \to Set) (pp : PTree\,(A \times A)) \to
  P\,(A \times A)\,(Pair^{\wedge}\,A\,A\,Q\,Q)\,pp \to
  P\,A\,Q\,(pnode\,pp) \big) \to} \\ \quad \mathsf{\forall (A : Set) (Q
  : A \to Set) (p : PTree\,A) \to PTree^{\wedge}\,A\,Q\,p \to
  P\,A\,Q\,p }\\
 \\
\mathsf{\forall (P : \forall (A : Set) \to (A \to Set) \to Bush\, A \to Set)
\to \big( \forall (A : Set)\,(Q : A \to Set) \to P\,A\,Q\,bnil \big) \to} \\ 
\quad\mathsf{\big( \forall (A : Set) (Q : A \to Set) (a : A) (bb :
  Bush\,(Bush\,A)) \to Q\,a \to
  P\,(Bush\,A)\,(P\,A\,Q)\,bb \to P\,A\,Q\,(bcons\,a\,bb)
  \big) \to} \\ 
\quad\mathsf{\forall (A : Set) (Q : A \to Set) (b : Bush\,A) \to
  Bush^{\wedge}\,A\,Q\,b \to P\,A\,Q\,b } 
\end{array}\]}

\vspace*{-0.1in}

\caption{Deep induction rules for perfect trees and
  bushes}\label{fig:ptree-and-bush}
\end{adjustbox}
\end{figure*}

The deep induction rule for perfect trees similarly uses polymorphic
predicates but otherwise follows the familiar pattern. It is given by
the first expression in Figure~\ref{fig:ptree-and-bush}. There,
$\mathsf{Pair^{\wedge} : \forall (A\; B: Set) \to (A \to Set) \to (B
  \to Set)\; \to\;}$ $\mathsf{A \times B \to}$ $\mathsf{Set}$ lifts
predicates $\mathsf{Q_A}$ on data of type $\mathsf{A}$ and
$\mathsf{Q_B}$ on data of type $\mathsf{B}$ to a predicate on pairs of
type $\mathsf{A \times B}$ in such a way that
$\mathsf{Pair^{\wedge}\,A\,B\,Q_A\,Q_B\,(a,b) = Q_A\,a \times
  Q_B\,b}$. Similarly, $\mathsf{PTree^{\wedge}}$ $\mathsf{\,:
  \,\forall (A : Set) \to (A \to Set) \to PTree\,A \to Set}$ lifts a
predicate $\mathsf{Q}$ on data of type $\mathsf{A}$ to a predicate on
data of type $\mathsf{PTree\,A}$ asserting that $\mathsf{Q}$ holds for
every element of type $\mathsf{A}$ contained in its perfect tree
argument.  A general definition of liftings for a robust class of
GADTs including all those appearing in the literature is given in
Section 5.
%\looseness=-1

Using deep induction we can extend structural induction to {\em truly
  nested types}, i.e., to nested types whose recursive occurrences
appear below themselves. The quintessential example of such a type is
that of bushes\footnote{To define truly nested types in Agda we must
  use the $\mathsf{NO\_POSITIVITY\_CHECK}$ flag, and to define
  functions over them we must use the $\mathsf{TERMINATING}$ flag.
  (Similar flags are required in Coq.) Although as programmers we know
  from the metatheory in~\cite{jp19} that $\mathsf{Bush}$ is
  well-defined and the functions we define over them terminate, the
  flags are necessary because Agda fails to infer these
  facts. Analogous comments apply at several places
  below.}\cite{bm98}:
\begin{equation*}\label{eq:bush}
\begin{array}{l}
\mathsf{data\ Bush : Set \to Set\ where}\\
\mathsf{\;\;bnil\,\,\,\,\,\; :\, \forall \{A : Set\} \to Bush\,A}\\
\mathsf{\;\;bcons\, :\, \forall \{A : Set\} \to A \to Bush\,(Bush\,A) \to Bush\,A} 
\end{array}
\end{equation*}
Even defining a structural induction rule for bushes requires that we
be able to lift the rule's polymorphic predicate argument to
$\mathsf{Bush}$ itself. This observation was, in fact, the original
motivation for the development of deep induction in~\cite{jp20}. The
deep induction rule for bushes is given by the second expression in
Figure~\ref{fig:ptree-and-bush}, where
\[\mathsf{Bush^{\wedge} :
  \forall (A : Set) \to (A \to Set) \to Bush\,A \to Set}\]
is the following lifting of a predicate $\mathsf{Q}$ on data of type
$\mathsf{A}$ to a predicate on data of type $\mathsf{Bush\,A}$
asserting that $\mathsf{Q}$ holds for every element of type
$\mathsf{A}$ contained in its argument bush:
\begin{equation}
\begin{array}{l}
\mathsf{Bush^{\wedge}\,A\,Q\,bnil} \hspace*{0.49in} = \; \mathsf{\top} \\
\mathsf{Bush^{\wedge}\,A\,Q\,(bcons\,a\,bb)} \; = \;\\
\hspace*{0.3in}\mathsf{Q\,a \times Bush^{\wedge}\,(Bush\,A)\,(Bush^{\wedge}\,A\,Q)\,bb} 
\end{array}
\end{equation}

We note that, as for ADTs, the structural induction rule for any
(truly) nested type can be obtained as the special case of its deep
induction rule in which the custom predicates are taken to be
constantly $\mathsf{\top}$-valued predicates. This instantiation
ensures that the resulting induction rule only inspects the top-level
structure of its argument, rather than the contents of that structure,
which is exactly what structural induction should do.\looseness=-1
%A concrete example of such a derivation is given in
%Section~\ref{sec:ind-equal}.

Under some circumstances deep induction can be mimicked by
hand-threading applications of structural induction through the layers
of data comprising a deep data type. But this is not the case if,
e.g., one or more of the custom predicates in the data type's deep
induction rule is not the characteristic function of an inductive data
type.

\section{(Deep) GADTs}\label{sec:GADTs}

While a data constructor for a nested type can take {\em as arguments}
data whose types involve instances of that type at indices other than
the one being defined, its return type must still be at the (variable)
type instance being defined. For example, each of $\mathsf{pleaf}$ and
$\mathsf{pnode}$ returns an element of type $\mathsf{PTree\,A}$
regardless of the instances of $\mathsf{PTree}$ appearing in the types
of its arguments. GADTs relax this restriction, allowing their data
constructors both to take as arguments \emph{and return as results}
data whose types involve instances other than the one being
defined. That is, GADTs' constructors' return type instances can, like
that of $\mathsf{pair}$ in~\eqref{eq:seq}, be structured.  For every
GADT in this paper, we require that the instance of the return type
for each of its data constructors is a polynomial in that
constructor's type arguments.\looseness=-1

GADTs are used in precisely those situations in which different
behaviors at different instances of data types are desired. This is
achieved by allowing the programmer to give the type signatures of the
GADT's data constructors independently, and then using pattern
matching to force the desired type refinement. For example, the {\em
  equality} GADT
\begin{equation}\label{eq:equal}
\begin{array}{l}
\mathsf{data\ Equal : Set \to Set \to Set\ where}\\
\mathsf{\;\;refl :\, \forall \{A : Set\} \to Equal\,A\,A}
\end{array}
\end{equation}
is parameterized by two type indices, but it is only possible to
construct data elements of type $\mathsf{Equal\,A\,B}$ if $\mathsf{A}$
and $\mathsf{B}$ are instantiated at the same type. If the types
$\mathsf{A}$ and $\mathsf{B}$ are syntactically identical then the
type $\mathsf{Equal\,A\,B}$ contains the single data element
$\mathsf{refl}$. It contains no data elements otherwise.

\begin{figure*}[t]

\begin{adjustbox}{varwidth=7.3in, max width=7.1in, margin=-0.0in 0in
      -0.2in 0in, center} 

{\small
\[\begin{array}{l}
\mathsf{data\ LType : Set \to Set\ where}\\
\mathsf{\;\;bool :\, \forall \{A : Set\} \to \forall (B : Set) \to Equal\,A\,Bool
  \to LType\,A}\\ 
\mathsf{\;\;arr\;\;\; :\, \forall \{A : Set\} \to \forall (B\,C : Set) \to
  Equal\,A\,(B \to C) \to LType\,B \to LType\,C \to LType\,A}\\ 
  \mathsf{\;\;list\,\;\; :\,  \forall \{A : Set\} \to\forall (B : Set) \to
    Equal\,A\,(\List\,B) \to LType\,B \to LType\,A}\\
  \\
\mathsf{data\ LTerm : Set \to Set\ where}\\
\mathsf{\;\;var\,\,\,:\,  \forall \{A : Set\} \to String \to LType\,A \to
  LTerm\,A} \\  
\mathsf{\;\;abs\,\, :\,  \forall \{A : Set\} \to \forall (B\,C : Set) \to
  Equal\,A\,(B \to C) \to String \to LType\,B \to LTerm\,C \to
  LTerm\,A}\\ 
  \mathsf{\;\;app\; :\,  \forall \{A : Set\} \to \forall (B : Set) \to
    LTerm (B \to A) \to LTerm\,B \to LTerm\,A} \\ 
  \mathsf{\;\;list\;\,\, :\,  \forall \{A : Set\} \to \forall (B : Set) \to
    Equal\,A\,(List\,B) \to List\,(LTerm\,B) \to LTerm\,A} 
\end{array}\]}

\vspace*{-0.1in}

\caption{The $\mathsf{LType}$ and $\mathsf{LTerm}$ data
  types}\label{fig:type-and-term}
\end{adjustbox}
\end{figure*}

The importance of the equality GADT lies in the fact that we can
understand other GADTs in terms of it. For example, the GADT
$\mathsf{Seq}$ from~\eqref{eq:seq} comprises constant sequences of
data of any type $\mathsf{A}$ and sequences obtained by pairing the
data in two already existing sequences. This GADT can be rewritten as
its Henry Ford encoding~\cite{ch03,hin03,mcb99,sjsv09,sp04}, which
makes critical use of the equality GADT, as follows:
\begin{equation}\label{eq:eq_seq}
\begin{array}{l}
\mathsf{data\ Seq : Set \to Set\ where}\\
\mathsf{\;\;const :\, \forall \{A : Set\} \to A \to Seq\,A}\\ 
\mathsf{\;\;pair\,\;\; :\, \forall \{A : Set\} \to \forall (B\,C : Set) \to
  Equal\,A\,(B \times C) \to}\\
\mathsf{\hspace*{0.7in}Seq\,B \to Seq\,C \to Seq\,A}\\ 
\end{array}
\end{equation}
Here, the requirement that $\mathsf{pair}$ produce data at an instance
of $\mathsf{Seq}$ that is a product type is replaced with the
requirement that $\mathsf{pair}$ produce data at an instance of
$\mathsf{Seq}$ that is \emph{equal} to a product type. As we will see
in Section~\ref{sec:deep-ind-GADTs}, this encoding in terms of the
equality GADT is key to deriving deep induction rules for GADTs.

Although $\mathsf{Seq}$ does not at first glance appear to be a deep
GADT, when written as its Henry Ford encoding, it, like all GADTs, can
be regarded as ``deep over $\mathsf{Equal}$''.  By contrast, the
%Neither $\mathsf{Equal}$ nor $\mathsf{Seq}$ is a deep GADT, but the
GADT $\mathsf{LTerm}$ in Figure~\ref{fig:type-and-term}, which is
inspired by~\cite{cis194}, is {\em inherently deep}.  It encodes terms
of a simply typed lambda calculus. More robust variations on
$\mathsf{LTerm}$ are, of course, possible, but this variation is rich
enough to illustrate all essential aspects of deep GADTs --- and
later, in Section~\ref{sec:ind-lam}, their deep induction rules ---
while still being small enough to ensure clarity of exposition.

Types are either booleans, arrow types, or list types. They are
represented by the Henry Ford GADT $\mathsf{LType}$ in
Figure~\ref{fig:type-and-term}.  Terms are either variables,
abstractions, applications, or lists of terms. They are similarly
represented by the Henry Ford GADT $\mathsf{LTerm}$.  The type
parameter for $\mathsf{LTerm}$ tracks the types of simply typed lambda
calculus terms. For example, $\mathsf{LTerm\,A}$ is the type of simply
typed lambda terms of type $\mathsf{A}$. Variables are tagged with
their types by the data constructors $\mathsf{var}$ and
$\mathsf{abs}$, whose $\mathsf{LType}$ arguments ensure that their
type tags are legal types. This ensures that all lambda terms produced
by $\mathsf{var}$, $\mathsf{abs}$, $\mathsf{app}$, and $\mathsf{list}$
are well-typed.  We will revisit these GADTs in
Sections~\ref{sec:deep-ind-GADTs} and~\ref{sec:app}.

\section{(Deep) Induction for GADTs}\label{sec:deep-ind-GADTs}

The equality constraints engendered by GADTs' data constructors makes
deriving (deep) induction rules for them more involved than for ADTs
and other nested types. Nevertheless, we show in this section how to
do so. We first illustrate the key components of our approach by
deriving deep induction rules for the three specific GADTs introduced
in Section~\ref{sec:GADTs}. Then, in Section~\ref{sec:framework}, we
abstract these to a general framework that can be applied to any GADT
that is not a truly nested GADT. As hinted above, the predicate
lifting for the equality GADT plays a central role in deriving both
structural and deep induction rules for more general GADTs.
 
\subsection{(Deep) Induction for $\mathsf{Equal}$}\label{sec:ind-equal}

To define the (deep) induction rule for any GADT $\mathsf{G}$ we first
need to define a predicate lifting that maps a predicate on a type
$\mathsf{A}$ to a predicate on $\mathsf{G\,A}$. Such a predicate
lifting
\[\begin{array}{l}
\mathsf{Equal^{\wedge} : \forall (A\,B : Set) \to (A \to Set)
  \to (B \to Set) \to}\\
\mathsf{\hspace*{0.7in}Equal\,A\,B \to Set}
\end{array}\]
for $\mathsf{Equal}$ is defined by
\[\mathsf{Equal^{\wedge}\,A\,A\,Q\,Q'\,refl = \forall (a :
  A) \to Equal\,(Q\,a)(Q'\,a)}\]
It does exactly what we expect: it takes two predicates on the same
type as input and is inhabited iff they are extensionally equal.
Next, we need to associate with each data constructor $\mathsf{c}$ of
$\mathsf{G}$ an {\em induction hypothesis} asserting that, if the
custom predicate arguments to a predicate $\mathsf{P}$ on $\mathsf{G}$
can be lifted to $\mathsf{G}$ itself, then $\mathsf{c}$ {\em respects}
$\mathsf{P}$, i.e., $\mathsf{c}$ constructs data satisfying the
instance of $\mathsf{P}$ at those custom predicates. The following
induction hypothesis $\mathsf{dIndRefl}$ is thus associated with the
$\mathsf{refl}$ constructor for $\mathsf{Equal}$:
\begin{equation*}\label{eq:ind-refl}
\begin{array}{l}
\mathsf{\lambda (P : \forall (A\,B : Set) \to (A \to Set) \to (B \to
  Set) \to}\\
\mathsf{\hspace*{1.8in}Equal\,A\,B \to Set) \to} \\ 
\quad\mathsf{\forall (C : Set) (Q\, Q' : C \to Set) \to
  Equal^{\wedge}\,C\,C\,Q\,Q'\,refl \to}\\
\mathsf{\hspace*{2in}P\,C\,C\,Q\,Q'\,refl} 
\end{array}
\end{equation*}
The deep induction rule for $\mathsf{G}$ now states that, if all of
$\mathsf{G}$'s data constructors respect a predicate $\mathsf{P}$,
then $\mathsf{P}$ is satisfied by every element of $\mathsf{G}$ to
which the custom predicate arguments to $\mathsf{P}$ can be
successfully lifted.  The deep induction rule for $\mathsf{Equal}$ is
thus
\begin{equation}\label{eq:ind-equal}
\begin{array}{l}
\mathsf{\forall (P : \forall (A\,B : Set) \to (A \to Set) \to (B \to
  Set) \to}\\
\mathsf{\hspace*{1in}Equal\,A\,B \to Set) \to dIndRefl\,P \to}\\ \quad 
\mathsf{\forall (A\,B : Set) (Q_A : A \to Set) (Q_B : B \to Set)(e:
  Equal\,A\,B)}\\
\mathsf{\hspace*{1in}\to Equal^{\wedge}\,A\,B\,Q_A\,Q_B\,e \to
  P\,A\,B\,Q_A\,Q_B\,e}\\ 
\mathsf{\hspace*{2in} }
\end{array}
\end{equation}

To prove that this rule is sound we must provide a witness
$\mathsf{dIndEqual}$ inhabiting the type in~\eqref{eq:ind-equal}.  By
pattern matching, we need only consider the case where $\mathsf{A} =
\mathsf{B}$ and $\mathsf{e} = \mathsf{refl}$, so we can define
$\mathsf{dIndEqual}$ by
\[\mathsf{dIndEqual\;P\;crefl\;A\;A\;Q_A\;Q_A'\;refl\;liftE =
  crefl\;A\;Q_A\;Q_A'\,liftE}\]
To recover $\mathsf{Equal}$'s structural induction rule
\begin{equation}\label{eq:sind-equal}
\begin{array}{l}
  \mathsf{\forall (Q : \forall (A\,B : Set)
  \to Equal\,A\,B \to Set) \to}\\
\mathsf{\hspace*{0.15in}\big( \forall (C : Set) \to
  Q\,C\,C\,refl \big) \to}\\
\mathsf{\hspace*{0.15in}\forall (A\,B : Set) (e: Equal\,A\,B) \to
  Q\,A\,B\,e}
\end{array}
\end{equation}
we define a term $\mathsf{indEqual}$ of the type
in~\eqref{eq:sind-equal} by $\mathsf{indEqual\;Q}$
$\mathsf{srefl\;A\;B\;refl = dIndEqual\;P\;srefl'\;A\;B\;
  K^A_\top\;K^B_\top\;refl\,sliftE}$.\\
Here,
\[\begin{array}{l}
\mathsf{P : \forall
  (A\,B : Set) \to (A \to Set) \to (B \to Set) \to}\\
\mathsf{\hspace*{2in}Equal\,A\,B} \mathsf{\to Set}
\end{array}\]
is defined by $\mathsf{P\;A\;B\;Q_A\;Q_B\;e =}$ $\mathsf{Q\;A\;B\;e}$,
$\mathsf{K^A_\top}$ and $\mathsf{K^B_\top}$ are the constantly
$\mathsf{\top}$-valued predicates on $\mathsf{A}$ and $\mathsf{B}$,
respectively, $\mathsf{sliftE :}$ $\mathsf{Equal^{\wedge}\;A\;B\;
  K^A_\top\;K^B_\top\;refl}$ is defined by \[\mathsf{sliftE\,a =}
\mathsf{refl : Equal\, \top\,\top}\] for every $\mathsf{a : A}$, and
\begin{equation*}
\begin{array}{l}
\mathsf{srefl' : \forall (C\,: Set) (Q_c\,Q'_c\, : C \to Set) \to}\\
\mathsf{\hspace*{0.75in} Equal^{\wedge}\,C\,C\,Q_c\,Q'_c\,refl \to \,Q\,C\,C\,refl}
\end{array}
\end{equation*}
is
defined by $\mathsf{srefl'\,C\,Q_c\,Q'_c\,liftE'\,=srefl\, C}$.  The
structural induction rule for any GADT $\mathsf{G}$ that is not truly
nested can similarly be recovered from its deep induction rule by
instantiating every custom predicate by the appropriate constantly
$\mathsf{\top}$-valued predicate.

\subsection{(Deep) Induction for $\mathsf{Seq}$}\label{sec:ind-seq}

\begin{figure*}[t]

\begin{adjustbox}{varwidth=7.3in, max width=7.1in, margin=-0.0in 0in
      -0.2in 0in, center} 

{\small
\[\begin{array}{lll}
\mathsf{Seq^{\wedge}\,A\,Q_A\,(const\,a)} & = & \mathsf{Q_A\,a}\\
\mathsf{Seq^{\wedge}\,A\,Q_A\,(pair\,B\,C\,e\,s_B\,s_C)}
&=&\mathsf{\exists [Q_B] \exists [Q_C]\, Equal^{\wedge}\,A\, (B
  \times C)\, Q_A\, (Pair^\wedge\,B\,C\, Q_B \, Q_C) \, e \times
  Seq^{\wedge}\,B\,Q_B\,s_B \times Seq^{\wedge}\,C\,Q_C\,s_C}\\
& & \\
\mathsf{dIndConst} & = & \mathsf{\lambda (P : \forall (A : Set) \to (A
  \to Set) \to Seq\,A \to Set) \to}\\
 & & \mathsf{\hspace*{0.15in}
\forall (A : Set) (Q_A : A \to Set) (a : A) \to Q_A\,a \to
P\,A\,Q_A\,(const\,a)}\\
& & \\
\mathsf{dIndPair} & = & \mathsf{\lambda (P : \forall (A : Set) \to (A
  \to Set) \to Seq\,A \to Set)} \to \\ 
 & & \quad \mathsf{\forall (A\,B\,C : Set) (Q_A : A \to Set) (Q_B : B
  \to Set) (Q_C : C \to Set)}\\
& & \mathsf{\hspace*{1.5in}(s_B : Seq\,B) (s_C : Seq\,C) (e :
  Equal\,A\,(B \times C)) \to} \\ 
& & \quad \mathsf{Equal^{\wedge} A\, (B \times C)\, Q_A\,
  (Pair^{\wedge}\,B\,C\,Q_B\,Q_C)\, e \to P\,B\,Q_B\,s_B \to}\\
& & \mathsf{\hspace*{1.5in}  P\,C\,Q_C\,s_C \to P\, A\, Q_A\, (
  pair\,B\,C\,e\,s_B\,s_C )} 
\end{array}\]

\vspace*{0.1in}

\[\begin{array}{l}
\hspace*{-2in}\mathsf{\forall (P : \forall (A : Set) \to (A \to Set) \to Seq\,A \to
  Set)} \mathsf{\to dIndConst\,P \to dIndPair\,P \to} \\ \hspace*{-2in}\quad
\mathsf{\forall (A : Set)(Q_A : A \to Set)(s_A : Seq\,A) \to
  Seq^{\wedge}\,A\,Q_A\,s_A \to P\,A\,Q_A\,s_A}
\end{array}\]}

\vspace*{-0.1in}

\caption{Deep induction rule for $\mathsf{Seq}$}\label{fig:seq}
\end{adjustbox}
\end{figure*}

To derive the deep induction rule for the GADT $\mathsf{Seq}$ we use
its Henry Ford encoding from~\eqref{eq:eq_seq}.  We first define its
predicate lifting
\[\mathsf{Seq^\wedge : \forall (A : Set) \to (A \to
  Set) \to Seq\,A \to Set}\] as in Figure~\ref{fig:seq}. There,
$\mathsf{a : A}$, $\mathsf{Q_B : B \to Set}$, $\mathsf{Q_C : C \to
  Set}$, $\mathsf{e : Equal\,A\,(B \times C)}$, $\mathsf{s_B :
  Seq\,B}$, $\mathsf{s_C : Seq\,C}$, and $\mathsf{\exists [x]\, F
  \,x}$ is syntactic sugar for the type of dependent pairs
$\mathsf{(x,b)}$, where $\mathsf{x : A}$, $\mathsf{b : F\, x}$, and
$\mathsf{F : A \to Set}$.  The lifting $\mathsf{Seq^\wedge}$ is
derived as in Section~\ref{sec:framework}. Next, let
$\mathsf{dIndConst}$ and $\mathsf{dIndPair}$ be the induction
hypotheses associated with the constructors $\mathsf{const}$ and
$\mathsf{pair}$, respectively. These are given in Figure~\ref{fig:seq}
as well. Then the deep induction rule for $\mathsf{Seq}$ is given in
the last two lines of Figure~\ref{fig:seq}.

To prove that this rule is sound we provide a witness
$\mathsf{dIndSeq}$ inhabiting the type in the last two lines of
Figure~\ref{fig:seq} by\looseness=-1
\[\begin{array}{ll}
& \!\!\mathsf{dIndSeq\;P\;cconst\;cpair\;A\;Q_A\;(const\,a)\;liftA}\\
= & \!\!\mathsf{cconst\;A}$ $\mathsf{Q_A\;a\;liftA}
\end{array}\]
and
\[\begin{array}{ll}
 &\!\!\mathsf{dIndSeq\,P\,cconst\,cpair\,A\,Q_A\,(pair\,B\,C\,e\,s_B\,s_C)}\\
 & \hspace*{1.4in}\mathsf{(Q_B, Q_C, liftE, liftB, liftC)}\\
= &\!\! \mathsf{cpair\,A\,B\,C\,Q_A\,Q_B\,Q_C\,s_B\,s_C\,e\,liftE\,
  p_B\,p_C}
\end{array}\]
In the first clause, $\mathsf{a : A}$, $\mathsf{Q_A : A \to Set}$, and
$\mathsf{liftA :}$ $\mathsf{Seq^{\wedge}\,A\,Q_A}$ $\mathsf{(const\,a)
  = Q_A\,a}$. In the second clause we also have
\[\begin{array}{lll}
\mathsf{Q_B} & : & \mathsf{B \to Set}\\
\mathsf{Q_C} & : & \mathsf{C \to Set}\\
\mathsf{e}  & : & \mathsf{Equal\,A\,(B \times C)}\\
\mathsf{s_B} & : & \mathsf{Seq\,B}\\
\mathsf{s_C} & : & \mathsf{Seq\,C}\\
\mathsf{liftE} & : & \mathsf{Equal^{\wedge}\,A\, (B \times C)\, Q_A\,
  (Pair^\wedge \,B\,C\,Q_B\,Q_C) \, e}\\
\mathsf{liftB} & : & \mathsf{Seq^{\wedge}\,B\,Q_B\,s_B}\\
\mathsf{liftC} & : & \mathsf{Seq^{\wedge}\,C\,Q_C\,s_C}
\end{array}\]
%$\mathsf{Q_B : B \to}$ $\mathsf{Set}$, $\mathsf{Q_C : C \to Set}$,
%$\mathsf{e \,:\, Equal\,A\,(B \times C)}$, \, $\mathsf{s_B :}$
%$\mathsf{Seq\,B}$, $\mathsf{s_C : Seq\,C}$,\, $\mathsf{liftE :
%Equal^{\wedge}\,A\,}$ $\mathsf{(B \times C)\, Q_A\, (Pair^\wedge
%\,B\,C\,Q_B}$ $\mathsf{Q_C) \, e}$, $\mathsf{liftB :
%Seq^{\wedge}\,B}$ $\mathsf{Q_B\,s_B}$, and $\mathsf{liftC}$ $\mathsf{
%: Seq^{\wedge}\,C\,Q_C\,s_C}$
Together these give that
\[\mathsf{(Q_B, Q_C, liftE,liftB, liftC)
\, :\,  Seq^{\wedge}\,A\,Q\,(pair\,B\,C\,e\,s_B\,s_C)}\]
We therefore have
\[\begin{array}{lllll}
\mathsf{p_B}\!\! & = & \!\!\mathsf{dIndSeq\,P
  \,cconst\,cpair\,B\,Q_B\,s_B\,liftB} \!\! & : & 
\!\!\mathsf{P\,B\,Q_B\,s_B}\\
\mathsf{p_C}\!\! & = & \!\!\mathsf{dIndSeq\,P\,
  cconst\,cpair\,C\,Q_C\,s_C\,liftC}\!\! & : &\!\! \mathsf{P\,C\,Q_C\,s_C} 
\end{array}\]

\subsection{(Deep) Induction for $\mathsf{LTerm}$}\label{sec:ind-lam} 

\begin{figure*}[t]

%  \begin{adjustbox}{varwidth=6.8in, max width=6.6in, margin=-0.1in 0in
%      -0.2in 0in, fbox, center} 

  \begin{adjustbox}{varwidth=7.3in, max width=7.1in, margin=-0.0in 0in
      -0.2in 0in, center} 

  {\small
\[\begin{array}{lll}
\mathsf{LType^{\wedge}\,A\,Q_A\,(bool\,B\,e)} & = &\mathsf{\exists
  [Q_B]\, Equal^{\wedge}\, A\, B\, Q_A\, K^{Bool}_{\top} \,e}\\
\mathsf{LType^{\wedge}\,A\,Q_A\,(arr\, B\, C\, e\, T_B\, T_C)}
&=&\mathsf{\exists [Q_B] \,\exists [Q_c]\, Equal^{\wedge}\,A\,
  (B \to C)\, Q_A\, (Arr^{\wedge} \, B\, C\, Q_B \, Q_C) \, e \times
  \, LType^{\wedge}\,B\,Q_B\,T_B \times LType^{\wedge}\,C\,Q_C\,T_C}\\
\mathsf{LType^{\wedge}\,A\,Q_A\,(list\, B\, e\, T_B)} & = &
\mathsf{\exists [Q_B]\, Equal^{\wedge}\,A\, (List\, B)\, Q_A\,
  (List^{\wedge} \, B\, Q_B) \, e \times LType^{\wedge}\,B\,Q_B\,T_B}\\[1ex]
\mathsf{LTerm^{\wedge}\,A\,Q_A\,(var\,s\,T_A)} & = &
\mathsf{LType^{\wedge}\, A\, Q_A\, T_A}\\
\mathsf{LTerm^{\wedge}\,A\,Q_A\, (abs \,B \,C \,e \,s \,T_B \,t_C)} &
= & \mathsf{\exists [Q_B]\,\exists [Q_C]\, Equal^{\wedge} \, A\, (B \to
  C)\, Q_A\, (Arr^{\wedge} \, B\, C\, Q_B \, Q_C)\, e \times \,
  LType^{\wedge}\, B\, Q_B\, T_B \times \, LTerm^{\wedge}\, C\, Q_C\,
  t_C }\\
\mathsf{LTerm^{\wedge}\,A\,Q_A\, (app\, B\, t_{BA}\, t_B)} & = &
\mathsf{\exists [Q_B]\, LTerm^{\wedge}\, (B \to A)\, (Arr^{\wedge} \,
  B\, A\, Q_B \, Q_A)\, t_{BA} \times LTerm^{\wedge}\, B\, Q_B\,
  t_B}\\
\mathsf{LTerm^{\wedge}\,A\,Q_A\, (list\, B\, e\, ts)} & = &
\mathsf{\exists [Q_B]\, Equal^{\wedge} \, A\, (List\,B)\, Q_A\,
  (List^{\wedge} \, B\, Q_B) \, e \times List^{\wedge}\, (LTerm\,B) \,
  (LTerm^{\wedge} \, B\, Q_B) \, ts}
\end{array}   \]}

\vspace*{-0.1in}

\caption{Predicate liftings for $\mathsf{LType}$ and
  $\mathsf{LTerm}$}\label{fig:liftings} \vspace*{0.1in} 
\end{adjustbox}
\end{figure*}

To derive the deep induction rule for the GADT $\mathsf{LTerm}$ we use
its Henry Ford encoding from Figure~\ref{fig:type-and-term}.
We first define the predicate lifting
\[\begin{array}{l}
\mathsf{Arr^{\wedge} : \forall (A\, B : Set) \to (A \to Set) \to (B
  \to Set) \to}\\
\mathsf{\hspace*{2.2in}(A \to B) \to Set}
\end{array}\] for arrow types following the general framework in
Section~\ref{sec:framework}, since 
arrow types appear in $\mathsf{LType}$ and $\mathsf{LTerm}$.  It is
given by \[\mathsf{Arr^{\wedge}\, A\, B\, Q_A\, Q_B\, f = \forall (a :
  A) \to Q_A\,a \to Q_B\, (f\,a)}\] The predicate liftings
\[\mathsf{LType^{\wedge} : \forall (A : Set) \to (A \to Set) \to
  LType\,A \to Set}\] for $\mathsf{LType}$ and
\[\mathsf{LTerm^{\wedge}
  : \forall (A : Set) \to (A \to Set) \to LTerm\,A \to Set}\] for
$\mathsf{LTerm}$ are defined in Figure~\ref{fig:liftings} following
the general framework in Section~\ref{sec:framework}.  There,
\[
\begin{array}{lll}
\mathsf{s} & : & \mathsf{String} \\
\mathsf{Q_A} & : & \mathsf{A \to Set} \\
\mathsf{Q_B} & : & \mathsf{B \to Set} \\
\mathsf{Q_C} & : & \mathsf{C \to Set} \\
\mathsf{T_A} & : & \mathsf{LType\, A} \\
\mathsf{T_B} & : & \mathsf{LType\, B} \\
\mathsf{T_C} & : & \mathsf{LType\, C} \\
\mathsf{t_B} & : & \mathsf{LTerm \, B} \\
\mathsf{t_C} & : & \mathsf{LTerm \, C} \\
\mathsf{t_{BA}} & : & \mathsf{LTerm \, (B \to A)} \\
\mathsf{ts} & : & \mathsf{List\, (LTerm\, B)}
\end{array}
\]
and $\mathsf{K^{Bool}_{\top}}$ is the constantly
$\mathsf{\top}$-valued predicate on $\mathsf{Bool}$ and
$\mathsf{List^\wedge}$ is the predicate lifting for lists
from~\eqref{eq:rose}.  Also,
\[
\begin{array}{llll}
\mathsf{e} & : & \mathsf{Equal\,A\,Bool} & \text{in the first clause,} \\
\mathsf{e} & : & \mathsf{Equal\, A\, (B \to C)} & \text{in the second and fifth clauses,} \\
\mathsf{e} & : & \mathsf{Equal\, A\, (List\, B)} & \text{in the third clause,} \\
\mathsf{e} & : & \text{Equal\, A\, (List \,B)} & \text{in the seventh clause.}
\end{array}
\]
%$\mathsf{s : String}$, $\mathsf{Q_A : A \to}$ $\mathsf{Set}$,
%$\mathsf{Q_B : B \to Set}$, $\mathsf{Q_C : C \to Set}$,
%$\mathsf{K^{Bool}_{\top}}$ is the constantly $\mathsf{\top}$-valued
%predicate on $\mathsf{Bool}$, $\mathsf{T_A : LType\, A}$, $\mathsf{T_B
 % : LType \,B}$, $\mathsf{T_C :}$ $\mathsf{LType \,C}$, $\mathsf{t_B :
 % LTerm \, B}$, $\mathsf{t_C : LTerm \, C}$, and $\mathsf{t_{BA} :
% LTerm \, (B \to A)}$.  Also, $\mathsf{e : Equal\,A\,Bool}$ in the
%first clause, $\mathsf{e : Equal\, A\, (B \to C)}$ in the second and
%fifth, $\mathsf{e : Equal\, A\, (List\, B)}$ in the third,
%$\mathsf{e : Equal \, A \, (B \to C)}$ in the fifth,
%and $\mathsf{e : Equal\, A\, (List \,B)}$,
%$\mathsf{ts : List\, (LTerm B)}$, and $\mathsf{List^\wedge}$ is the
%predicate lifting for lists from~\eqref{eq:rose} in the seventh.

With these liftings in hand we can define the induction hypotheses
$\mathsf{dIndVar}$, $\mathsf{dIndAbs}$, $\mathsf{dIndApp}$, and
$\mathsf{dIndList}$ associated with $\mathsf{LTerms}$'s data
constructors. These are given in Figure~\ref{fig:ind-hyps-lterm}.
The deep induction rule for $\mathsf{LTerm}$ is thus
\begin{equation}\label{eq:ind-lam}
\begin{array}{l}
\mathsf{\forall (P : \forall (A : Set) \to (A \to Set) \to LTerm\,A
  \to Set) \to}\\
\quad\mathsf{dIndVar\,P \to dIndAbs\,P \to dIndApp\,P \to
  dIndList\,P \to}\\
\quad\quad \mathsf{\forall (A : Set)(Q_A : A \to
  Set)(t_A : LTerm\,A) \to}\\
\quad\quad\quad\mathsf{LTerm^{\wedge}\,A\,Q_A\,t_A \to
  P\,A\,Q_A\,t_A}
\end{array}
\end{equation}

\begin{figure*}[t]

\begin{adjustbox}{varwidth=7.3in, max width=7.1in, margin=-0.0in 0in
      -0.2in 0in, center} 

{\small
\[\begin{array}{lll}
\mathsf{dIndVar} & = & \mathsf{\lambda (P : \forall (A : Set) \to (A \to Set) \to
  LTerm\,A  \to Set) \to}\\
& &   \quad\mathsf{\forall (A : Set) (Q_A : A \to Set) (s : String) (T_A :
  LType\, A) \to LType^{\wedge} \, A\, Q_A\, T_A \to P \, A\, Q_A\,
 (var \; s\, T_A)}\\[1ex]
\mathsf{dIndAbs} & = & \mathsf{\lambda (P : \forall (A : Set) \to (A \to Set) \to
  LTerm\,A \to Set) \to} \\ 
& &   \quad\mathsf{
  \forall (A\,B\,C: Set) (Q_A : A \to Set)  (Q_B : B \to Set) (Q_C : C
  \to Set) (e : Equal\, A\, (B \to C)) (s : String) \to } \\ 
& &   \quad\mathsf{(T_B : LType\, B) \to (t_C : LTerm\, C)
  \to Equal^{\wedge}\,A\,(B \to C)\, Q_A \, (Arr^{\wedge} \, B\, C\,
  Q_B \, Q_C) \, e \to  } \\
& &   \quad\mathsf{
  LType^{\wedge}\, B\, Q_B\, T_B
  \to P\, C\, Q_C\, t_C\, 
  \to P \, A\, Q_A\, (abs \,B \,C \, e \,s \,T_B \, t_C)}\\
 \\
\mathsf{dIndApp} & = &  \mathsf{\lambda (P : \forall (A : Set) \to (A \to Set) \to LTerm\,A
    \to Set)\to} \\ 
& &  \quad \mathsf{
  \forall (A \,B : Set) (Q_A : A \to Set)  (Q_B : B \to Set) 
   (t_{BA} : LTerm\, (B \to A)) (t_B : LTerm\, B) \to} \\
& &   \quad \mathsf{
  P\, (B \to A)\, (Arr^{\wedge} \, B\, A\, Q_B \, Q_A) \, t_{BA} \, 
  \to P\, B\, Q_B\, t_B\, 
  \to P \, A\, Q_A\, (app \,B \,t_{BA} \, t_B) }\\[1ex]
\mathsf{dIndList} & = &   \mathsf{\lambda (P : \forall (A : Set) \to (A \to Set) \to LTerm\,A
    \to Set) \to} \\ 
& &   \quad \mathsf{
  \forall (A \,B : Set) (Q_A : A \to Set)  (Q_B : B \to Set) 
    (e : Equal\, A\, (List\, B)) (ts : List\, (LTerm\, B)) \to} \\ 
& &   \quad \mathsf{
    Equal^{\wedge}\, A\, (List\,B)\, Q_A\, (List^{\wedge}\, B\, Q_B)\, e 
  \to List^{\wedge}\, (LTerm\,B) (P\, B\, Q_B)\, ts
  \to P \, A\, Q_A\, (list \,B \,e \, ts) }
\end{array}\]}

\vspace*{-0.1in}

\caption{Induction hypotheses for $\mathsf{LTerm}$}\label{fig:ind-hyps-lterm}
\end{adjustbox}
\end{figure*}

\begin{figure*}[t]

    \begin{adjustbox}{varwidth=7.3in, max width=7.1in, margin=-0.0in 0in
      -0.2in 0in, center} 
  
%  \begin{adjustbox}{varwidth=6.8in, max width=6.6in, margin=-0.1in 0in
%      -0.2in 0in, fbox, center} 
{\small
\[\begin{array}{lll}
\mathsf{dIndLTerm \, P\, cvar \, cabs\, capp\, clist \, A\, Q_A\,
  (var\;s\,T_A) \, liftA} & = & \mathsf{cvar \, A\, Q_A\, s\, T_A\,
  liftA}\\ 
\mathsf{dIndLTerm \, P\, cvar \, cabs\, capp\, clist \, A\, Q_A\,
  (abs \,B \,C \,e \,s \,T_B \, t_C) \, (Q_B , Q_C , liftE,
  lift_{T_B}, lift_{t_C})} & = & \mathsf{cabs\,A\,B\,C\, Q_A\,
  Q_B\, Q_C\, e\, s\, T_B\, t_C\, liftE\, lift_{T_B}\, p_C}\\
\mathsf{dIndLTerm \, P\, cvar \, cabs\, capp\, clist \, A\, Q_A\,
    (app \,B \,\,t_{BA} \, t_B)\, (Q_B , lift_{t_{BA}}, lift_{t_B})} &
= & \mathsf{capp\,A\,B\,Q_A\, Q_B\, t_{BA}\, t_B\, p_{BA} \, p_B}\\
  \mathsf{dIndLTerm \, P\, cvar \, cabs\, capp\, clist \, A\, Q_A\,
    (list \,B \,e \, ts) \, (Q_B , liftE', lift_{List})} & = & 
  \mathsf{clist \,A\,B\,Q_A\, Q_B\, e\, ts\, liftE'\, p_{List} }
\end{array}\]}
\mbox{where}
{\small
\[\begin{array}{lll}
\mathsf{p_C} & = & \mathsf{dIndLTerm\,P\,cvar\,cabs \,capp \,clist\,
  C\, Q_C\, t_C\, lift_{t_C} : P \, C\, Q_C \, t_C }\\
\mathsf{p_B} & = & \mathsf{dIndLTerm\,P\,cvar\,cabs \,capp \,clist\,
  B\, Q_B\, t_B\, lift_{t_B} : P \, B\, Q_B \, t_B }\\
\mathsf{p_{BA}} & = & \mathsf{dIndLTerm\,P\,cvar\,cabs \,capp
  \,clist\, (B \to A)\,(Arr^{\wedge} \, B\, A\, Q_B \, Q_A) \,
  t_{BA}\, lift_{t_{BA}} : P \, (B \to A)\, (Arr^{\wedge} \, B\, A\,
  Q_B \, Q_A) \, t_{BA}}\\ 
\mathsf{p_{List}} & = &\mathsf{liftListMap \, (LTerm\, B) \,
  (LTerm^{\wedge} \, B \, Q_B)\, (P\,B\,Q_B)\, p_{ts} \, ts\,
  lift_{List} : List^{\wedge}\, (LTerm\,B) \, (P\,B\,Q_B) \, ts}\\
\mathsf{p_{ts}} & = & \mathsf{dIndLTerm\, P\, cvar\, cabs\, capp\,
  clist\, B\, Q_B : PredMap\,(LTerm\,B) \,(LTerm^{\wedge}\, B\, Q_B)
  \, (P\,B\,Q_B)}
\end{array}\]}

\vspace*{-0.1in}

\caption{$\mathsf{dIndLTerm}$}\label{fig:dindlterm} \vspace*{0.1in} 
\end{adjustbox}
\end{figure*}

To prove that this rule is sound we define a witness
$\mathsf{dIndLTerm}$ inhabiting the type in~\eqref{eq:ind-lam} as in
Figure~\ref{fig:dindlterm}. There,

\[
\begin{array}{lll}
\mathsf{s} & : & \mathsf{String} \\
\mathsf{Q_A} & : & \mathsf{A \to Set} \\
\mathsf{Q_B} & : & \mathsf{B \to Set} \\
\mathsf{Q_C} & : & \mathsf{C \to Set} \\
\mathsf{T_A} & : & \mathsf{LType\, A} \\
\mathsf{T_B} & : & \mathsf{LType\, B} \\
\mathsf{t_B} & : & \mathsf{LTerm \, B} \\
\mathsf{t_C} & : & \mathsf{LTerm \, C} \\
\mathsf{t_{BA}} & : & \mathsf{LTerm \, (B \to A)} \\
\mathsf{ts} & : & \mathsf{List\, (LTerm\, B)}\\
\mathsf{liftA} & : & \mathsf{LTerm^{\wedge}\, A\, Q_A\, (var\;s\,T_A) = LType^{\wedge}\,A\,Q_A\,T_A} \\
\mathsf{liftE} & : & \mathsf{Equal^{\wedge}\, A\, (B \to C)\, Q_A\, (Arr^{\wedge} \, B\, C\, Q_B \, Q_C) \, e} \\
\mathsf{lift_{T_B}} & : & \mathsf{LType^{\wedge} \, B\, Q_B\, T_B} \\
\mathsf{lift_{t_C}} & : & \mathsf{LTerm^{\wedge} \, C\, Q_C\, t_C} \\
\mathsf{lift_{t_{BA}}} & : & \mathsf{LTerm^{\wedge} \, (B \to A)\, (Arr^{\wedge} \, B\, A\, Q_B \, Q_A)\, t_{BA}}
\end{array}\]

\[\begin{array}{lll}
\mathsf{lift_{t_B}} & : & \mathsf{LTerm^{\wedge} \, B\, Q_B\, t_B} \\
\mathsf{liftE'} & : & \mathsf{Equal^{\wedge}\, A\, (List\,B)\, Q_A\, (List^{\wedge}\, B\, Q_B)\, e} \\
\mathsf{lift_{List}} & : & \mathsf{List^{\wedge} \, (LTerm\, B) \,
  (LTerm^{\wedge}\, B\, Q_B) \, ts}
\end{array}
\]
%$\mathsf{s : String}$, $\mathsf{Q_A
%  : A \to Set}$, $\mathsf{Q_B : B \to Set}$, $\mathsf{Q_C : C \to
%  Set}$, $\mathsf{T_A : LType\,A}$, $\mathsf{T_B}$ $\mathsf{:
%  LType\,B}$, $\mathsf{t_B : LTerm\,B}$, $\mathsf{t_C : LTerm\,C}$,
%$\mathsf{t_{BA} : LTerm\,(B \to A)}$, $\mathsf{liftA :
%  LTerm^{\wedge}\, A\, Q_A\, (var\;s\,T_A) =
%  LType^{\wedge}\,A\,Q_A\,T_A}$, $\mathsf{liftE :}$
%$\mathsf{Equal^{\wedge}}$ $\mathsf{A\, (B \to C)\, Q_A\, (Arr^{\wedge}
%  \, B\, C\, Q_B \, Q_C) \, e}$, $\mathsf{lift_{T_B}: LType^{\wedge}
%  \, B\, Q_B\, T_B}$, $\mathsf{lift_{t_C}}$ $\mathsf{: LTerm^{\wedge}
%  \, C\, Q_C\, T_C}$, $\mathsf{lift_{t_{BA}}: LTerm^{\wedge} \, (B \to
%  A)\, (Arr^{\wedge} \, B\, A\, Q_B \, Q_A)}$ $\mathsf{t_{BA}}$,
%$\mathsf{lift_{t_B}: LTerm^{\wedge} \, B\, Q_B\, t_B}$,
%$\mathsf{liftE' : Equal^{\wedge}\, A\, (List\,B)\, Q_A}$ and
%$\mathsf{(List^{\wedge}\, B\, Q_B)\, e}$, and $\mathsf{lift_{List}:
%  List^{\wedge} \, (LTerm\, B) \, (LTerm^{\wedge}\, B\, Q_B) \, ts}$.
Moreover, in the definition of $\mathsf{p_{ts}}$, \[\mathsf{PredMap :
  \forall\, (A : Set) \to (A \to Set) \to (A \to
  Set) \to Set }\] is the type constructor producing the type of
morphisms between predicates defined by\looseness=-1
\[\mathsf{PredMap \,A\, Q\,Q'\,
  = \forall\, (a : A) \to Q\,a \to Q'\,a}\] and
\[\begin{array}{l}
\mathsf{liftListMap : \forall\, (A : Set) \to (Q \, Q' : A \to Set)
  \to}\\\
\hspace*{0.8in}\mathsf{PredMap\,A\,Q\,Q' \to}\\
\hspace*{1in}\mathsf{PredMap\,(List\,A)
  \,(List^{\wedge}\, A\, Q)\, (List^{\wedge}\, A\, Q')}
\end{array}\]
which takes a morphism $\mathsf{f}$ of predicates and produces a
morphism of lifted predicates, is defined by
\[\mathsf{liftListMap\, A\, Q\, Q'\, m\, nil\, tt = tt}\]
(since $\mathsf{x : List^{\wedge}\,
  A\, Q\, nil}$ must necessarily be the sole inhabitant $\mathsf{tt}$
of $\mathsf{\top}$), and by
\[\begin{array}{ll}
 & \mathsf{liftListMap\, A\, Q\, Q'\, m\,
  (cons\, a\, l')\, (y, x')}\\
= & \mathsf{(m\,a\,y, \,liftListMap\, A\,
  Q\, Q'\, m\, l'\, x')}
\end{array}\] (since $\mathsf{x : List^{\wedge}\, A\,
  Q\, (cons\, a\, l')}$ must be of the form $\mathsf{x = (y, x')}$
where $\mathsf{y : Q\,a}$ and $\mathsf{x' : List^{\wedge}\, A\, Q\,
  l'}$).

\vspace*{-0.05in}

\section{The General Framework}\label{sec:framework}

We can generalize the approach in Section~\ref{sec:deep-ind-GADTs} to
a general framework for deriving deep induction rules for GADTs that
are not truly nested GADTs. We will treat GADTs of the form

\vspace*{-0.05in}

\begin{equation}\label{eq:gadts}
\begin{array}{l}
  \mathsf{data\ G : Set^\alpha
    \to Set\ where}\\
\mathsf{\;\;\;\;\;\;\;\;c\, :\, \forall \{\ol{B : Set}\} \to F\,G\,\ol{B} \to G (\ol{K\,\ol{B}})}
\end{array}
\end{equation}

\noindent
For brevity and clarity we indicate only one constructor $\mathsf{c}$
in~\eqref{eq:gadts}, even though a GADT can have any finite number of
them, each with a type of the same form as
$\mathsf{c}$'s. In~\eqref{eq:gadts}, $\mathsf{F}$ and each
$\mathsf{K}$ in $\ol{\mathsf{K}}$ are type constructors with
signatures $\mathsf{(Set^{\alpha} \to Set) \to Set^{\beta} \to Set}$
and $\mathsf{Set^{\beta} \to Set}$, respectively. If $\mathsf{T}$ is a
type constructor with signature $\mathsf{Set^{\gamma} \to Set}$ then
$\mathsf{T}$ has {\em arity} $\mathsf{\gamma}$.  The
overline notation denotes a finite list whose length is exactly the
arity of the type constructor being applied to it. The number of type
constructors in $\ol{\mathsf{K}}$ (resp., $\ol{\mathsf{B}}$) is thus
$\alpha$ (resp., $\beta$). In addition, the type constructor
$\mathsf{F}$ must be constructed inductively according to the
following grammar:\label{grammar}
\[\begin{array}{lll}
\mathsf{F\,G\,\ol{B}} & := &
\mathsf{F_1\,G\,\ol{B} \times F_2\,G\,\ol{B} \ \vert\ F_1\,G\,\ol{B} +
  F_2\,G\,\ol{B}}\\
& & \!\!\!\!\mathsf{\vert\ F_1\,\ol{B} \to F_2\,G\,\ol{B}
\ \vert\ G\,(\ol{F_1\,\ol{B}}) \ \vert\ H\,\ol{B} \ \vert\ H\,
(\ol{F_1\,G\,\ol{B}}) }
\end{array}\]
This grammar is subject to the following restrictions. In the third
clause the type constructor $\mathsf{F_1}$ does not contain
$\mathsf{G}$. In the fourth clause, none of the $\mathsf{\alpha}$-many
type constructors in $\mathsf{\ol{F_1}}$ contains $\mathsf{G}$. This
prevents nesting, which would make it impossible to give an induction
rule for $\mathsf{G}$; see Section~\ref{sec:GADT-nested} below. In the
fifth and sixth clauses, $\mathsf{H : Set^\gamma \to Set}$ is the
syntactic reflection of some functor, and thus has an associated map
function. It is worth noting that the fifth clause subsumes the cases
in which $\mathsf{F\,G\,\ol{B}}$ is a closed type or one of the
$\mathsf{B_i}$, and that $\mathsf{H}$ can be the data type constructor
for any (truly) nested type. From the map function for $\mathsf{H}$ we
can also construct a map function
\begin{equation}\label{eq:hliftmap}
\begin{array}{lll}
\mathsf{H^\wedge Map} & : & \mathsf{\forall (\ol{A : Set}) (\ol{Q\;Q'
    : A \to Set}) \to}\\
& & \mathsf{\ol{PredMap\,A\,Q\,Q'} \to}\\
& & \mathsf{PredMap\,(H\,\ol{A})\,(H^{\wedge}\,\ol{A}\,\ol{Q})\, 
(H^{\wedge}\,\ol{A}\,\ol{Q'})}
\end{array}
\end{equation}
for $\mathsf{H^{\wedge}}$. A concrete way to define $\mathsf{H^\wedge
  Map}$ is by induction on the structure of the type $\mathsf{H}$, but
we omit such details since they are not essential to the present
discussion. A further requirement that applies to all of the type
constructors appearing in the right-hand side of the above grammar,
including those in $\ol{\mathsf{K}}$, is that they must all admit
predicate liftings. This is not an overly restrictive condition,
though: all
%types constructed from sums, products, arrow types and
%type application admit predicate liftings, and so do
GADTs constructed from the above grammar admit predicate liftings.
(The fact that the domain of an arrow type is independent of
$\mathsf{G}$ is crucial for this.) In particular, the lifting for each
type constructor $\mathsf{H}$ is constructed using its map function.
%In fact, we have seen such liftings for
%products and type application in Section~\ref{sec:deep-ind-GADTs}.
A concrete way to define more general predicate liftings is, again, by
induction on the structure of the types in a suitable calculus; this
will ensure that the liftings satisfy the crucial property needed to
derive deep induction rules, namely that of distributing over the type
constructors. We do not give a general definition of predicate
liftings here, since that would require us to first design a full type
calculus, which is beyond the scope of the present paper. We can
however, define liftings for the type constructor $\mathsf{F}$ defined
by the grammar on page~\pageref{grammar} by
\begin{itemize}
\item $\mathsf{F\,G\,\ol{B} = F_1\,G\,\ol{B} \times F_2\,G\,\ol{B}}$ then
\[\begin{array}{ll}
& \!\!\mathsf{F^{\wedge}\,G\,\ol{B}\,P\,\ol{Q_B}}\\
\quad\quad= & \!\!\mathsf{Pair^{\wedge}\,(F_1\,G\,\ol{B})\, (F_2\,G\,\ol{B})\,
  (F_1^{\wedge}\,G\,\ol{B}\,P\,\ol{Q_B})\,
  (F_2^{\wedge}\,G\,\ol{B}\,P\,\ol{Q_B})}
  \end{array}\]
\item $\mathsf{F\,G\,\ol{B} = F_1\,G\,\ol{B} + F_2\,G\,\ol{B}}$ then
\[\begin{array}{ll}
& \!\!\mathsf{F^{\wedge}\,G\,\ol{B}\,P\,\ol{Q_B}}\\
\quad\quad= & \!\!\mathsf{Pair^{\wedge}\,(F_1\,G\,\ol{B})\, (F_2\,G\,\ol{B})\,
  (F_1^{\wedge}\,G\,\ol{B}\,P\,\ol{Q_B})\,
  (F_2^{\wedge}\,G\,\ol{B}\,P\,\ol{Q_B})}
  \end{array}\]
\item If $\mathsf{F\,G\,\ol{B} = F_1\,\ol{B} \to F_2\,G\,\ol{B}}$ then
  \[\begin{array}{ll}
  & \!\!\mathsf{F^{\wedge}\,G\,\ol{B}\,P\,\ol{Q_B}\,x}\\
  \quad\quad= & \!\!\mathsf{\forall (z :
  F_1\,\ol{B}) \to F_1^{\wedge}\,\ol{B}\,\ol{Q_B}\,z \to
  F_2^{\wedge}\,G\,\ol{B}\,P\,\ol{Q_B}\,(x\,z)}
  \end{array}\]
\item If $\mathsf{F\,G\,\ol{B} = G\,(F_1\,\ol{B})}$ and $\mathsf{F_1}$
  does not contain $\mathsf{G}$, then
  \[\mathsf{F^{\wedge}\,G\,\ol{B}\,P\,\ol{Q_B} = P \,(F_1\,\ol{B})\,
    (F_1^{\wedge}\,\ol{B}\,\ol{Q_B})}\] for all $\mathsf{P : \forall (A
    : Set)\to}$ $\mathsf{(A \to Set) \to G\,A \to Set}$.
\item If $\mathsf{F\,G\,\ol{B} = H\,\ol{B}}$ and $\mathsf{H}$ does not
  contain $\mathsf{G}$, then
  \[\mathsf{F^{\wedge}\,G\,\ol{B}\,P\,\ol{Q_B} = H^\wedge\,\ol{Q_B}}\]
  for all $\mathsf{P : \forall (A : Set)\to(A \to Set) \to G\,A \to}$
  $\mathsf{ Set}$.
\item If $\mathsf{F\,G\,\ol{B} = H\, (\ol{F_k\,G\,\ol{B}})}$ and
  $\mathsf{H}$ does not contain $\mathsf{G}$, then
  \[\mathsf{F^{\wedge}\,G\,\ol{B}\,P\,\ol{Q_B} = H^{\wedge}\,
  (\ol{F_k\,G\,\ol{B}})\,
    (\ol{F_k^{\wedge}\,G\,\ol{B}\,P\,\ol{Q_B}})}\] for all $\mathsf{P
    : \forall (A : Set)\to(A \to Set) \to G\,A \to Set}$.
\end{itemize}

We assume in the development below that $\mathsf{G}$ is a unary type
constructor, i.e., that $\alpha = 1$ in~\eqref{eq:gadts}. Extending
the argument to GADTs of arbitrary arity presents no difficulty other
than heavier notation. In this case the type of $\mathsf{G}$'s single
data constructor $\mathsf{c}$ can be rewritten as
\[\mathsf{c : \forall
  (\ol{B : Set}) \to Equal\,A\,(K\,\ol{B})\to F\,G\,\ol{B} \to G\,A}\]
The predicate lifting $\mathsf{G^{\wedge} : \forall (A : Set) \to (A
  \to Set) \to G\,A \to}$\\ $\mathsf{Set}$ for $\mathsf{G}$ is therefore
\[\begin{array}{l}
\mathsf{G^{\wedge}\,A\,Q_A\,(c\,\ol{B}\,e\,x)
  =}\\
\mathsf{\hspace*{0.15in}\exists [\ol{Q_B}]\,
Equal^{\wedge}\,A\,(K\,\ol{B})\,Q_A\,(K^{\wedge}\,\ol{B}\,\ol{Q_B})\,e
\times F^{\wedge}\,G\,\ol{B}\,G^{\wedge}\,{\ol{Q_B}}\,x}
\end{array}\]
where $\mathsf{Q_A : A \to Set}$, $\ol{\mathsf{Q_B : B \to Set}}$,
$\mathsf{e : Equal\,A\,(K\,\ol{B})}$, and $\mathsf{x :}$\\
$\mathsf{F\,G\,\ol{B}}$.
If we have predicate liftings
\[\begin{array}{l}\mathsf{F^{\wedge}} : \mathsf{\forall
  (G : Set^{\alpha} \to Set) (\ol{B : Set}) \to}\\
\mathsf{\hspace*{0.3in}(\forall (A : Set) \to
  (A \to Set) \to G\,A \to Set) \to}\\
\mathsf{\hspace*{0.3in}(\ol{B \to Set}) \to F\,G\,\ol{B}
  \to Set}
\end{array}\] for $\mathsf{F}$ and \[\mathsf{K^{\wedge}} :
\mathsf{\forall (\ol{B : Set}) \to (\ol{B \to Set}) \to K\,\ol{B} \to
  Set}\]
for $\mathsf{K}$, then the induction hypothesis $\mathsf{dIndC}$
associated with $\mathsf{c}$ is
\[\begin{array}{l}
\mathsf{dIndC = \lambda (P : \forall (A : Set) \to (A \to Set) \to
  G\,A \to Set) \to}\\
\quad \mathsf{\forall (A : Set)\, (\ol{B : Set})\, (Q_A : A \to
  Set)\,(\ol{Q_B : B \to Set})}\\ 
\quad \quad \mathsf{(e : Equal\,A\,(K\,\ol{B}))\, (x : F\,G\,\ol{B})
  \to}\\
\quad \quad \quad
\mathsf{Equal^{\wedge}\,A\,(K\,\ol{B})\,Q_A\,(K^{\wedge}\,\ol{B}\,\ol{Q_B})\,e 
  \to}\\
\quad\quad\quad\quad\mathsf{F^{\wedge}\,G\,\ol{B}\,P\,\ol{Q_B}\,x \to
  P\,A\,Q_A\,(c\,\ol{B}\,e\,x)} 
\end{array}\]
and the induction rule for $\mathsf{G}$ is
\begin{equation}\label{eq:gen-ind-rule}
\begin{array}{l}
  \mathsf{\forall (P : \forall (A : Set) \to (A \to Set) \to G\,A \to
  Set) \to}\\
\quad \mathsf{dIndC\,P \to \forall (A : Set)(Q_A : A \to Set)(y : G\,A)
  \to}\\
\quad\quad\mathsf{G^{\wedge}\,A\,Q_A\,y \to P\,A\,Q_A\,y}
\end{array}
\end{equation}

To prove that this rule is sound we define a witness
$\mathsf{dIndG}$ inhabiting this type by
\[\begin{array}{ll}
 & \mathsf{dIndG\,P\,cc\,A\,Q_A\,(c\,\ol{B}\,e\,x)\,(\ol{Q_B}, liftE, liftF)}\\
= & \mathsf{cc\,A\,\ol{B}\,Q_A\,\ol{Q_B}\,e\,x\,liftE\,(p\,x\,liftF)}
\end{array}\]
Here,
\[\begin{array}{lll}
\mathsf{cc} & : & \mathsf{dIndC\,P}\\
\mathsf{e} & : & \mathsf{Equal\,A\,(K\,\ol{B})}\\
\mathsf{x} & : & \mathsf{F\,G\,\ol{B}}\\
\mathsf{Q_A} & : & \mathsf{A \to Set}\\
\mathsf{liftE} & : & \mathsf{Equal^{\wedge}\,A\,(K\,\ol{B})\,Q_A\,
  (K^{\wedge}\,\ol{B}\,\ol{Q_B})\,e}\\
\mathsf{liftF} & : &
\mathsf{F^{\wedge}\,G\,\ol{B}\,G^{\wedge}\,{\ol{Q_B}}\,x}
\end{array}\]
and $\ol{\mathsf{Q_B : B \to Set}}$,
so
\[\mathsf{(\ol{Q_B}, liftE, liftF) : G^{\wedge}\,A\,Q_A (c\,\ol{B}\,
  e\,x)}\] as expected.  Finally, the morphism of predicates
\[\mathsf{p : PredMap\,(F\,G\,\ol{B})
  (F^{\wedge}\,G\,\ol{B}\,G^{\wedge}\,\ol{Q_B}) 
  (F^{\wedge}\,G\,\ol{B}\,P\,\ol{Q_B})}\] is defined by structural
induction on $\mathsf{F}$ as follows:

\begin{itemize}
\item 
If $\mathsf{F\,G\,\ol{B} = F_1\,G\,\ol{B} \times F_2\,G\,\ol{B}}$ then
\[\begin{array}{ll}
& \!\!\mathsf{F^{\wedge}\,G\,\ol{B}\,P\,\ol{Q_B}}\\
\quad\quad= & \!\!\mathsf{Pair^{\wedge}\,(F_1\,G\,\ol{B})\, (F_2\,G\,\ol{B})%\,
  (F_1^{\wedge}\,G\,\ol{B}\,P\,\ol{Q_B})\,
  (F_2^{\wedge}\,G\,\ol{B}\,P\,\ol{Q_B})}
  \end{array}\] The
induction hypothesis ensures
%{\color{red} liftings $\mathsf{F_1}$ and $\mathsf{F_2}$ (with
%  types) and}
  morphisms of predicates
  \[\quad\quad\;\;\mathsf{p_1
  : PredMap\,(F_1\,G\,\ol{B})\,(F_1^{\wedge}\,G\,\ol{B}\,G^{\wedge}\,\ol{Q_BQ})
  (F_1^{\wedge}\,G\,\ol{B}\,P\,\ol{Q_B})}\] and \[\quad\quad\mathsf{p_2 :
  PredMap\,(F_2\,G\,\ol{B})\,(F_2^{\wedge}\,G\,\ol{B}\,G^{\wedge}\,\ol{Q_B})
  (F_2^{\wedge}\,G\,\ol{B}\,P\,\ol{Q_B})}\] For $\mathsf{x_1 :
  F_1\,G\,\ol{B}}$, $\mathsf{liftF_1 :
  F_1^{\wedge}\,G\,\ol{B}\,G^{\wedge}\,\ol{Q_B}\,x_1}$, $\mathsf{x_2 :
  F_2\,G\,\ol{B}}$ and $\mathsf{liftF_2}$ $\mathsf{:
  F_2^{\wedge}\,G\,\ol{B}\,G^{\wedge}\,\ol{Q_B}\,x_2}$ we then define
\[\quad\mathsf{p\, (x_1, x_2)\, (liftF_1, liftF_2) =
  (p_1\,x_1\,liftF_1,\, p_2\,x_2\,liftF_2)}\]
\item The case $\mathsf{F\,G\,\ol{B} = F_1\,G\,\ol{B} +
  F_2\,G\,\ol{B}}$ is analogous.
\item If $\mathsf{F\,G\,\ol{B} = F_1\,\ol{B} \to F_2\,G\,\ol{B}}$ then
  \[\begin{array}{ll}
  & \!\!\mathsf{F^{\wedge}\,G\,\ol{B}\,P\,\ol{Q_B}\,x}\\
  \quad\quad= & \!\!\mathsf{\forall (z :
  F_1\,\ol{B}) \to F_1^{\wedge}\,\ol{B}\,\ol{Q_B}\,z \to
  F_2^{\wedge}\,G\,\ol{B}\,P\,\ol{Q_B}\,(x\,z)}
  \end{array}\] where $\mathsf{x :
  F\,G\,\ol{B}}$. The induction hypothesis ensures
%  {\color{red}
% liftings $\mathsf{F_1}$ and $\mathsf{F_2}$ (with types) and}
  a morphism of predicates
  \[\quad\quad\;\;\mathsf{p_2 : PredMap\,(F_2\,G\,\ol{B})\,
  (F_2^{\wedge}\,G\,\ol{B}\,G^{\wedge}\,\ol{Q_B})\,
  (F_2^{\wedge}\,G\,\ol{B}\,P\,\ol{Q_B})}\]  We therefore define
$\mathsf{p\,x\,liftF : F^{\wedge}\,G\,\ol{B}\,P\,\ol{Q_B}\,x}$, where
$\mathsf{liftF}$ $\mathsf{: F^{\wedge}\,G\,\ol{B}\,G^{\wedge}\,\ol{Q_B}\,x}$, to
be \[\mathsf{p\,x\,liftF\;z\;liftF_1 = p_2\, (x\,z)\,
  (liftF\,z\,liftF_1)}\] for $\mathsf{z : F_1\,\ol{B}}$ and
$\mathsf{liftF_1 : F_1^{\wedge}\,\ol{B}\,\ol{Q_B}\,z}$. Note that
$\mathsf{F_1}$ not containing $\mathsf{G}$ is a necessary restriction
since the proof relies on
$\mathsf{F^{\wedge}\,G\,\ol{B}\,G^{\wedge}\,\ol{Q_B}\,x}$ and
$\mathsf{F^{\wedge}\,G\,\ol{B}\,P\,\ol{Q_B}\,x}$ having the same
domain $\mathsf{F_1^{\wedge}\,\ol{B}\,\ol{Q_B}\,z}$.
\item If $\mathsf{F\,G\,\ol{B} = G\,(F_1\,\ol{B})}$ and $\mathsf{F_1}$
  does not contain $\mathsf{G}$, then
  \[\mathsf{F^{\wedge}\,G\,\ol{B}\,P\,\ol{Q_B} = P \,(F_1\,\ol{B})\,
    (F_1^{\wedge}\,\ol{B}\,\ol{Q_B})}\] for all $\mathsf{P : \forall (A
    : Set)\to}$ $\mathsf{(A \to Set) \to G\,A \to Set}$.  We then define
  $\mathsf{p = dIndG\,P}$
   $\mathsf{cc\,(F_1\,\ol{B})\,(F_1^{\wedge}\,\ol{B}\,\ol{Q_B})}$.
\item If $\mathsf{F\,G\,\ol{B} = H\,\ol{B}}$ and $\mathsf{H}$ does not
  contain $\mathsf{G}$, then
  \[\mathsf{F^{\wedge}\,G\,\ol{B}\,P\,\ol{Q_B} = H^\wedge\,\ol{Q_B}}\]
  for all $\mathsf{P : \forall (A : Set)\to(A \to Set) \to G\,A \to}$ $\mathsf{
    Set}$.  We therefore define \[\mathsf{p : PredMap\, (H\,\ol{B})\,
    (H^{\wedge}\,\ol{B}\,\ol{Q_B})\,(H^{\wedge}\,\ol{B}\,\ol{Q_B})}\]
  to be the identity morphism on predicates.
\item If $\mathsf{F\,G\,\ol{B} = H\, (\ol{F_k\,G\,\ol{B}})}$ and
  $\mathsf{H}$ does not contain $\mathsf{G}$, then
  \[\mathsf{F^{\wedge}\,G\,\ol{B}\,P\,\ol{Q_B} = H^{\wedge}\,
  (\ol{F_k\,G\,\ol{B}})\,
    (\ol{F_k^{\wedge}\,G\,\ol{B}\,P\,\ol{Q_B}})}\] for all $\mathsf{P
    : \forall (A : Set)\to(A \to Set) \to G\,A \to Set}$.  Since
  $\mathsf{H}$ is not a GADT, $\mathsf{H^\wedge}$ has a map function
  $\mathsf{H^\wedge Map}$ as in~\eqref{eq:hliftmap}.  The induction
  hypothesis ensures morphisms of
  predicates \[\quad\quad\;\;\ol{\mathsf{p_k :
      PredMap\,(F_k\,G\,\ol{B})
      (F_k^{\wedge}\,G\,\ol{B}\,G^{\wedge}\,\ol{Q_B})
      (F_k^{\wedge}\,G\,\ol{B}\,P\,\ol{Q_B})}}\] We therefore define
  \[\;\;\quad\quad\mathsf{p = H^\wedge Map\,(\ol{F_k\,G\,\ol{B}})\,
    (\ol{F_k^{\wedge}\,G\,\ol{B}\,G^{\wedge}\,\ol{Q_B}})
  \,(\ol{F_k^{\wedge}\,G\,\ol{B}\,P\,\ol{Q_B}})\,\ol{p_k}}\]
\end{itemize}

\vspace*{0.05in}

Observing that the above development essentially uses the equality
GADT and its predicate lifting in the discrete category of types to
extend the lifting in~\cite{jp20} --- now specialized to the same
category --- to GADTs, we have established the following theorem:

\begin{theorem}
A GADT $\mathsf{G}$ of the form in \eqref{eq:gadts} admits the deep
induction rule in~\eqref{eq:gen-ind-rule}.
\end{theorem}

\section{Truly Nested GADTs Need Not Admit Deep Induction
    Rules}\label{sec:GADT-nested}

%{\em Truly nested GADTs} are proper GADTs whose recursive occurrences
%appear below themselves.
In Sections~\ref{sec:deep-ind-GADTs} and~\ref{sec:framework} we
derived deep induction rules for GADTs that are not truly nested
GADTs. Since both (truly) nested types and GADTs without true nesting
admit deep induction rules, we might expect truly nested GADTs to
admit them as well. Unfortunately, however, the techniques developed
in the previous sections do not extend to truly nested GADTs. Indeed,
while the induction rule for a data type generally relies on (unary)
parametricity of the model interpreting it, deep induction for a truly
nested type or a truly nested GADT crucially relies on this
interpretation being functorial.  Whereas ADTs and nested types both
admit functorial parametric semantics, proper GADTs admit parametric
semantics but do not admit functorial semantics. In this section we
show how the techniques developed in this paper for deriving deep
induction rules go wrong for truly nested GADTs by analyzing the
following very simple example:\looseness=-1

\vspace*{-0.1in}

\begin{equation}\label{gadt-nested}
\begin{array}{l}
\mathsf{data\ G : Set \to Set\ where}\\
\mathsf{\;\;\;\;\;\;\;\;\;c :\,  \forall \{A : Set\} \to G\,(G\,A) \to G\,(A \times A)}
\end{array}
\end{equation}
We acknowledge that $\mathsf{G}$ is semantically equivalent to the
empty data type, and thus has a trivial (deep) induction principle. We
could, of course, consider a more realistic counterexample, but this
would only add notational overhead for no gain in conceptual
clarity. Indeed, we need only exhibit a single GADT whose deep
induction rule cannot be obtained using the techniques of this paper,
and for which more robust techniques will therefore be needed if deep
induction rules are to be derived for them.

To see this, we first rewrite the constructor $\mathsf{c}$'s type as
\[\mathsf{c : \forall\, (B : Set) \to Equal\,A\,(B \times B)
  \to G\,(G\,B) \to G\,A}\]
The predicate lifting
\[\mathsf{G^{\wedge} : \forall\, (A : Set) \to (A \to Set) \to G\,A \to
  Set}\] for $\mathsf{G}$ is therefore
\[\begin{array}{l}
\mathsf{G^{\wedge}\,A\,Q_A\,(c\,B\,e\,x) =}\\
\mathsf{\hspace*{0.15in}\exists\, [Q_B]\,
Equal^{\wedge}\,A\,(B \times
B)\,Q_A\,(Pair^{\wedge}\,B\,B\,Q_B\,Q_B)\,e}\\
\mathsf{\hspace*{1.8in}\times \;G^{\wedge}\,(G\,B)\,(G^{\wedge}\,B\,Q_B)\,x}
\end{array}\]
where $\mathsf{Q_A : A \to Set}$, $\mathsf{Q_B : B \to Set}$, $\mathsf{e
  : Equal\,A\,(B \times B)}$, and $\mathsf{x : G\,(G\,B)}$.
The induction hypothesis $\mathsf{dIndC}$ for $\mathsf{c}$ is
\[\begin{array}{l}
\mathsf{\lambda\, (P : \forall\, (A : Set) \to (A \to Set) \to G\,A
  \to Set)\to} \\ 
\quad\mathsf{ \forall\, (A\;B : Set)\, (Q_A : A \to Set)\, (Q_B : B
  \to Set)}\\
\hspace*{1in}\mathsf{(e : Equal\,A\,(B \times B))\, (x : G\,(G\,B))\to} \\ 
\quad\mathsf{Equal^{\wedge}\,A\,(B \times
  B)\,Q_A\,(Pair^{\wedge}\,B\,B\,Q_B\,Q_B)\,e \to}\\
\hspace*{1in}\mathsf{P\,(G\,B)\,(P\,B\,Q_B)\,x 
	\to P\,A\,Q_A\,(c\,B\, e\,x)} 
\end{array}\]
so the deep induction rule for $\mathsf{G}$ is
\[\begin{array}{l}
\mathsf{\forall\, (P : \forall\, (A : Set) \to (A \to Set) \to G\,A \to Set)
  \to}\\
\quad\mathsf{dIndC\,P \to \forall\, (A : Set)\, (Q : A \to Set)\, (y : G\,A)
  \to}\\
\quad\quad\mathsf{G^{\wedge}\,A\,Q\,y \to P\,A\,Q\,y}
\end{array}\] But if we now try to show
that this rule is sound by constructing a witness $\mathsf{dIndG}$
inhabiting this type we run into problems. We can define
\[\begin{array}{ll}
  & \mathsf{dIndG\,P\,cc\,A\,Q\,(c\,B\,e\,x)\,(Q', liftE, liftG)}\\
= & \mathsf{cc\,A\,B\,Q\,Q'\,e\,x\,liftE\,p}
\end{array}\]
where
\[\begin{array}{lll}
\mathsf{cc} & : & \mathsf{dIndC\,P}\\
\mathsf{Q} & : & \mathsf{A \to Set}\\
\mathsf{Q'} & : & \mathsf{B \to Set}\\
\mathsf{e} & : & \mathsf{Equal\,A\,(B \times B)}\\
\mathsf{x} & : & \mathsf{G\,(G\,B)}\\
\mathsf{liftG} & : & \mathsf{G^{\wedge}\,(G\,B)\, (G^{\wedge}
  B\,Q')\,x}\\
\mathsf{liftE} & : & \mathsf{Equal^{\wedge}\,A\,(B \times 
  B)\,Q\,(Pair^{\wedge}\,B\,B\,Q'\,Q')\,e}
\end{array}\]
but we still need to
define $\mathsf{p : P\,(G\,B)\,(P\,B\,Q')\,x}$.  For this we can use
the induction rule and let
\[\mathsf{p = dIndG\,P\,cc\,(G\,B)\,(P\,B\,Q')\,x\,q}\] but we still
need to define \[\mathsf{q : G^{\wedge}\,(G\,B)\,(P\,B\,Q')\,x}\]  If
we had the map function
\[\begin{array}{l}
\mathsf{G^\wedge Map : \forall\, (A : Set)\,
  (Q\;Q' : A \to Set) \to}\\
\hspace*{0.7in}\mathsf{PredMap\,A\,Q\,Q' \to}\\
\hspace*{0.9in}\mathsf{PredMap\,(G\,A)\,(G^{\wedge}\,A\,Q)\,(G^{\wedge}\,A\,Q')} 
\end{array}\]
for $\mathsf{G^{\wedge}}$
then we could define
\[\mathsf{q = G^\wedge Map\,(G\,B)\,(G^{\wedge}\,B\,Q')\,(P\,B\,Q')\,
(dIndG\,P\,cc\,B\,Q')\,x\,liftG}\]  Unfortunately, however,
we cannot define $\mathsf{G^\wedge Map}$. Indeed, its definition would
have to be
\[\begin{array}{ll}
 & \!\!\mathsf{G^\wedge Map\,A\,Q\,Q'\,m\,(c\,B\,e\,x)\,(Q_B,
  liftE, liftG)}\\
\quad\quad= & \!\! \mathsf{(Q'_B, liftE', liftG')}
\end{array}\] for some
\[\begin{array}{lll}
\mathsf{Q'_B} & : & \mathsf{B \to Set}\\
\mathsf{liftE'} & :  & \mathsf{Equal^{\wedge}\,A\,(B \times
  B)\,Q'\,(Pair^{\wedge}\,B\,B\,Q'_B\,Q'_B)\,e}\\
\mathsf{liftG'} & : &
\mathsf{G^{\wedge}\,(G\,B)\,(G^{\wedge}\,B\,Q'_B)\,x}
\end{array}\]
where
\[\begin{array}{lll}
\mathsf{Q} & : & \mathsf{A \to Set}\\
\mathsf{Q'} & : & \mathsf{A \to Set}\\
\mathsf{Q_B} & : & \mathsf{B \to Set}\\
\mathsf{m} & : & \mathsf{PredMap\,A\,Q\,Q'}\\
\mathsf{e} & : & \mathsf{Equal\,A\,(B \times B)}\\
\mathsf{x} & : & \mathsf{G\,(G\,B)}\\
\mathsf{liftE} & : & \mathsf{Equal^{\wedge}\,A\,(B \times
  B)\,Q\,(Pair^{\wedge}\,B\,B\,Q_B\,Q_B)\,e}\\
\mathsf{liftG} & : &
\mathsf{G^{\wedge}\,(G\,B)\,(G^{\wedge}\,B\,Q_B)\,x}
\end{array}\]
That is, we would need to produce a proof $\mathsf{liftE'}$ of the
(extensional) equality of the predicates $\mathsf{Q'}$ and
$\mathsf{Pair^{\wedge}\,B\,B\,Q'_B\,Q'_B}$ from just a proof
$\mathsf{liftE}$ of the (extensional) equality of the predicates
$\mathsf{Q}$ and $\mathsf{Pair^{\wedge}\,B\,B\,Q_B\,Q_B}$ and a
morphism of predicates $\mathsf{m}$ from $\mathsf{Q}$ to
$\mathsf{Q_2'}$.
%a proof $\mathsf{liftE}$ of the (extensional) equality of the
%predicates $\mathsf{Q}$ and $\mathsf{Pair^{\wedge}\,B\,B\,Q_1\,Q_1}$
%and a morphism of predicates $\mathsf{m}$ from $\mathsf{Q}$ to
%$\mathsf{Q_2}$, and we need to use those data to deduce a proof of the
%(extensional) equality of the predicates $\mathsf{Q_2}$ and
%$\mathsf{Pair^{\wedge}\,B\,B\,Q_3\,Q_3}$, for some predicate
%$\mathsf{Q_3}$ on $\mathsf{B}$.
But this will not be possible in general: the facts that $\mathsf{Q}$
is equal to $\mathsf{Pair^{\wedge}\,B\,B\,Q_B\,Q_B}$ and that there is
a morphism of predicates $\mathsf{m}$ from $\mathsf{Q}$ to
$\mathsf{Q'}$ do not guarantee that there exists a predicate
$\mathsf{Q'_B}$ such that $\mathsf{Q'}$ is equal to
$\mathsf{Pair^{\wedge}\,B\,B\,Q'_B\,Q'_B}$.

At a deeper level, the fundamental issue is that the $\mathsf{Equal}$
type does not have functorial semantics~\cite{jg21}, so that having
morphisms $\mathsf{A \to A'}$ and $\mathsf{B \to B'}$ (for any type
$\mathsf{A, A', B}$ and $\mathsf{B'}$) and a proof that $\mathsf{A}$
is equal to $\mathsf{A'}$ does not provide a proof that $\mathsf{B}$
is equal to $\mathsf{B'}$. And not being able to define
\[\begin{array}{l}
\mathsf{Equal^\wedge Map : \forall (A\, B : Set)}\\
\hspace*{0.8in}\mathsf{(Q_A\, Q_A' : A \to Set)}\\
\hspace*{0.8in}\mathsf{(Q_B\, Q_B' : B \to Set) \to}\\
\hspace*{0.5in}\mathsf{PredMap\,A\,Q_A\,Q_A' \to}\\
\hspace*{0.5in}\mathsf{PredMap\,B\,Q_B\,Q_B' \to}\\
\hspace*{0.5in}\mathsf{PredMap\,(Equal\,A\,B)\,(Equal^\wedge
  A\,B\,Q_A\,Q_B)}\\
\hspace*{0.5in}\mathsf{(Equal^\wedge A\,B\,Q_A'\,Q_B')}
\end{array}\]
of course makes it unclear how to define $\mathsf{G^\wedge Map}$ for
more general $\mathsf{G}$.

\section{Case Study: Extracting Types of Lambda Terms}\label{sec:app}

In this section, we use deep induction for the $\mathsf{LTerm}$ GADT
from Figure~\ref{fig:type-and-term} to infer the type from a lambda
term. The following predicate either returns the type of its input
lambda term if that type can be inferred or indicates that the type
inference fails:
\begin{align*}
  &\mathsf{GetType : \forall \, (A : Set) \to LTerm\,A \to Set} \\
  &\mathsf{GetType \,A \,t = Maybe \, (LType \, A)}
\end{align*}
Of course, since $\mathsf{GetType}$ is (trivially) defined by
structural induction, we could perform type inference using
hand-threaded applications of structural induction as observed at the
end of Section~\ref{sec:ADTs-and-nesteds}. Nevertheless, the example
as given nicely illustrates deep induction.

\begin{figure*}[t]

\begin{adjustbox}{varwidth=7.3in, max width=7.1in, margin=-0.0in 0in
      -0.2in 0in, center} 

{\small
\[\begin{array}{lll}
\mathsf{cvar} & : & \mathsf{\forall\, (A : Set)\, (Q_A : A \to Set)\,
  (s : String)\, (T_A : LType\, A) \to LType^{\wedge} \, A\, Q_A\,
  T_A\, \to Maybe\, (LType\, A)}\\
 \mathsf{cabs} & : & \mathsf{\forall\, (A \, B\, C: Set)\, (Q_A : A
   \to Set)\, (Q_B : B \to Set)\, (Q_C : C \to Set)} \\ 
& &    \quad\mathsf{(e : Equal\, A\, (B \to C))\, (s : String)\, (T_B
   : LType\, B)\, (t_C : LTerm \, C)  \to} \\  
& & \quad\mathsf{Equal^{\wedge}\, A\, (B \to C)\, Q_A\,
   (Arr^{\wedge}\, B\, C\, Q_B\, Q_C)\, e  \to LType^{\wedge} \, B\,
   Q_B\, T_B  \to Maybe\, (LType\,C) \to Maybe\, (LType\, A)}\\
\mathsf{capp} & : & \mathsf{\forall\, (A \, B : Set)\, (Q_A : A \to
  Set)\, (Q_B : B \to Set)\,  (t_{BA}: LTerm\, (B \to A))\, (t_B :
  LTerm\, B)\to} \\  
& & \quad\mathsf{ Maybe\, (LType\, (B \to A))\to Maybe\, (LType\,
      B) \to Maybe\, (LType\, A)}\\
\mathsf{clist} & : & \mathsf{\forall\, (A \, B : Set)\, (Q_A : A \to
  Set)\, (Q_B : B \to Set)\, (e : Equal\, A\, (List\, B))\, (ts : 
  List\, (LTerm\, B)) \to}\\  
& & \quad\mathsf{Equal^{\wedge}\, A\, (List\,B) \, Q_A\,
  (List^{\wedge}\, B\, Q_B)\, e \to List^{\wedge}\, (LTerm\, B)\,
  (GetType\, B)\, ts \to Maybe\, (LType\, A)}
\end{array}\]}

\vspace*{-0.1in}

\caption{Applied induction hypotheses for
  $\mathsf{LTerm}$}\label{fig:ind-hyps-lterm-applied-to-P}
\end{adjustbox}
\end{figure*}

By construction every lambda term in $\mathsf{LTerm}$ is well-typed,
but that (necessarily unique) type cannot always be inferred.  The
predicate $\mathsf{GetType}$ uses the standard $\mathsf{Maybe}$ data
type to represent failure of type inference. It is defined
by:\looseness=-1

\vspace*{-0.1in}

\begin{equation}\label{eq:maybe}
\begin{array}{l}
\mathsf{data\ Maybe : Set \to Set\ where}\\
\mathsf{\;\;nothing :\,  \forall \{A : Set\} \to Maybe\,A}\\
\mathsf{\;\;just\;\;\;\;\;\;\; :\,  \forall \{A : Set\} \to A \to Maybe\,A}
\end{array}
\end{equation}

\noindent
We want to show that $\mathsf{GetType\,A\,t}$ is satisfied by every
element $\mathsf{t}$ in $\mathsf{LTerm\,A}$, i.e., we want to prove:
\[ \mathsf{getTypeProof : \forall \, (A : Set)\, (t : LTerm\,A) \to
   GetType \,A \,t}\] This property can be proved with deep
induction, which is used to apply the induction hypothesis to the
individual terms in the list of terms that the data constructor
$\mathsf{list}$ takes as an argument. Indeed, using the deep induction
rule $\mathsf{dIndLTerm}$ from Section~\ref{sec:ind-lam} we can define
$\mathsf{getTypeProof}$ by
\[\begin{array}{ll}
 & \mathsf{getTypeProof \,A \,t}\\
= & \mathsf{dIndLTerm\, P \, cvar\, cabs\, capp\, clist\, A\, K_\top\, t\,
  (LTerm^\wedge KT\, A\, t) }
\end{array}\] where $\mathsf{t : LTerm\,A}$,
$\mathsf{P}$ is the polymorphic predicate \[\mathsf{ \lambda \, (A:
  Set)\, (Q : A \to Set)\, (t : LTerm\,A)\, \to Maybe \, (LType \,
  A)}\] and $\mathsf{K_\top}$ is the constantly $\mathsf{\top}$-valued
predicate on $\mathsf{A}$, and \[\mathsf{LTerm^\wedge KT : \forall\,
  (A : Set)\, (t : LTerm A) \to LTerm^{\wedge}\, A\, K_\top\,t}\] is a
term, to be defined below, witnessing that $\mathsf{K_\top}$ can be
lifted to all terms. We also need the applications to $\mathsf{P}$ of
each of the induction hypotheses from Section~\ref{sec:ind-lam}. These
are given in Figure~\ref{fig:ind-hyps-lterm-applied-to-P}. In the first
clause, $\mathsf{cvar}$ returns $\mathsf{just\,T_A}$.  In the second
clause, $\mathsf{cabs}$ returns $\mathsf{nothing}$ if its final
argument is $\mathsf{nothing}$ and
\[\begin{array}{ll}
 & \!\!\mathsf{cabs\, A \,B \,C \,Q_A\, Q_B \,Q_C\, e \,s \,T_B\, t_C\, liftE\;
%  LLTy^\wedge B\,
  lift_{T_B}\; (just \,T_C)}\\
  = & \!\!\mathsf{just \, (arr\, B\, C\, e\, T_B \,T_C)}
\end{array}\] otherwise.
In the third clause,
\[\begin{array}{ll}
 & \!\!\mathsf{capp\, A \,B \, Q_A \, Q_B \, t_{BA}\, t_A \, (just \,
  (arr\, B \,A\, refl\, T_B\, T_A))\, mb}\\ 
= & \!\!\mathsf{just\, T_A}
\end{array}\]
and $\mathsf{capp}$ returns $\mathsf{nothing}$ otherwise.  In the
fourth clause, we must use $\mathsf{List^{\wedge}\, (LTerm\, B)\,
  (GetType\, B)\, ts}$ to extract the type of the head of
$\mathsf{ts}$ (from which we can deduce the type of the list). When
$\mathsf{ts} = \mathsf{nil}$ we define \[\mathsf{clist\, A\, B\, Q\,
  Q'\, e\, nil \, liftE\, lift_{ts} = nothing}\] where
$\mathsf{liftE : Equal^{\wedge}\, A\, (List\,B) \, Q\,
  (List^{\wedge}\, B\, Q')\, e}$, and $\mathsf{lift_{ts} :}$\\
$\mathsf{List^{\wedge}\, (LTerm\, B)\, (GetType\, B)\, ts}$.  When
$\mathsf{ts} = \mathsf{cons\,t\,ts'}$
% The interesting case is
%  $\mathsf{clist}$, in which we have to use the results of
%  $\mathsf{List^{\wedge}\, (LTerm\, B)\, (GetType\, B)\, ts}$ in order
%  to extract the type of one of the terms in the list.  To define
%  $\mathsf{clist}$ we pattern-match on the list of terms
%  $\mathsf{ts}$.  If $\mathsf{ts}$ is the empty list $\mathsf{nil}$,
%  we cannot extract a type, so we return $\mathsf{nothing}$:
%\[
%  \mathsf{clist\, A\, B\, Q\, Q'\, e\, nil \, liftE\, lift_{ts} = nothing}
%\]
%where $\mathsf{liftE : Equal^{\wedge}\, A\, (List\,B) \, Q\,
%  (List^{\wedge}\, B\, Q')\, e}$ and $\mathsf{lift_{ts} :
%  List^{\wedge}\, (LTerm\, B)\, (GetType\, B)\, ts}$.
%If $\mathsf{ts}$ is a non-empty list $\mathsf{cons\,t\,ts'}$, we
%pattern match on
%$\mathsf{list_{ts}}$ and use the result to construct
%the type we need.  The type of $\mathsf{list_{ts}}$ is
the type of $\mathsf{lift_{ts}}$ becomes
\[\begin{array}{ll}
& \!\!\mathsf{List^{\wedge}\, (LTerm\, B)\, (GetType\, B)\, (cons\, t}$
$\mathsf{ts')}\\
= & \!\!\mathsf{GetType\, B\, t \times List^{\wedge}\, (LTerm\, B)\,
  (GetType\, B)\, ts'}\\
= & \!\! \mathsf{Maybe\, (LType\, B) \times List^{\wedge}\, (LTerm\, B)\,
  (GetType\, B)\, ts'}
\end{array}\] We pattern match on the first component
of the pair to define\looseness=-1
%\[\begin{array}{lll}
%\mathsf{list_{ts}} & : &  
%  \mathsf{List^{\wedge}\, (LTerm\, B)\, (GetType\, B)\, (cons\, t\, ts')}\\
%& = & \mathsf{GetType\, B\, t \times List^{\wedge}\, (LTerm\, B)\,
%    (GetType\, B)\, ts'}\\ 
%&= & \mathsf{Maybe\, (LType\, B) \times List^{\wedge}\, (LTerm\, B)\,
%    (GetType\, B)\, ts'} 
%\end{array}\]
\begin{align*}
  &\mathsf{clist\, A\, B\, Q\, Q'\, e\, (cons\, t\, ts') \, liftE\,
    (nothing , lift_{ts'}) = nothing} \\ 
  &\mathsf{clist\, A\, B\, Q\, Q'\, e\, (cons\, t\, ts') \, liftE\,
    (just\, T' , lift_{ts'}) = just \, (list\, B\, e\, T')} 
\end{align*}
Here $\mathsf{e : Equal\, A\, (List\,B)}$, $\mathsf{T' : LType\, B}$,
and
\[\mathsf{lift_{ts'} : List^{\wedge}\, (LTerm\, B)\, (GetType\,
  B)\, ts'}\]

To finish defining $\mathsf{getTypeProof}$ we still need a proof
\[ \mathsf{LTerm^\wedge KT : \forall\, (A : Set)\, (t : LTerm\, A) \to
  LTerm^{\wedge}\, A\, K_\top\,t} \] Since $\mathsf{LTerm^\wedge}$ is
defined in terms of $\mathsf{LType^\wedge}$ and $\mathsf{Arr^\wedge}$,
and since $\mathsf{LType^\wedge}$ is also defined in terms of
$\mathsf{List^\wedge}$, we need analogous functions
$\mathsf{LType^\wedge KT}$, $\mathsf{Arr^\wedge KT}$ and
$\mathsf{List^\wedge KT}$, respectively, for each of these liftings as
well.  We only give the definition of $\mathsf{LTerm^\wedge KT}$ here
since $\mathsf{LType^\wedge KT}$, $\mathsf{Arr^\wedge KT}$, and
$\mathsf{List^\wedge KT}$ are defined analogously. We have:
\begin{itemize}
\item If $\mathsf{s : String}$ and $\mathsf{T : LType\,A}$ we define
  \[\mathsf{LTerm^\wedge KT\,A\,(var\,s\,T)} = \mathsf{LType^\wedge
  KT\,A\,T}\]
\item If $\mathsf{e : Equal\,A\,(B \to C)}$, $\mathsf{s : String}$,
  $\mathsf{T : LType\,B}$, and $\mathsf{t' :}$ $\mathsf{LTerm\,C}$ we
  need to define $\mathsf{LTerm^\wedge KT\,A\, (abs\, B \,C \, e \,s
    \,T \, t')}$ of type
\[\begin{array}{ll}
 & \!\!\mathsf{LTerm^{\wedge}\,A\,K_\top\, (abs \,B \,C \,e \,s \,T
  \,t')}\\
\quad\quad = & \!\!\mathsf{\exists [Q_B]\, [Q_C]\, Equal^{\wedge} \, A\, (B \to
  C)\, K_\top\, (Arr^{\wedge} \, B\, C\, Q_B \, Q_C)\, e }\\
 & \quad\quad\mathsf{\times \, LType^{\wedge}\, B\, Q_B\, T \times \,
  LTerm^{\wedge}\, C\, Q_C\,t' }
\end{array}\]
where $\mathsf{K_\top : A \to Set}$, $\mathsf{Q_B : B \to Set}$, and
$\mathsf{Q_C : C \to Set}$.  The only reasonable choice is to let both
$\mathsf{Q_B}$ and $\mathsf{Q_C}$ be $\mathsf{K_\top}$, which means we
need proofs of
$\mathsf{Equal^{\wedge} \, A\, (B \to C)}$ $\mathsf{K_\top\,
  (Arr^{\wedge} \, B\, C\, K_\top \, K_\top)\, e}$,
$\mathsf{LType^{\wedge}\, B\, K_\top\, T}$ and
$\mathsf{LTerm^{\wedge}\, C}$ $\mathsf{K_\top\, t'}$.  We take
$\mathsf{LType^\wedge KT\, B\, T}$ and $\mathsf{LTerm^\wedge KT\, C\,
  t'}$ for the latter two proofs. For the former we note that, since
we are working with proof-relevant predicates, the lifting
$\mathsf{Arr^{\wedge} \, B\, C\, K_\top \, K_\top}$ of
$\mathsf{K_\top}$ to arrow types is not identical to $\mathsf{K_\top}$
on arrow types but rather (extensionally) isomorphic.  We discuss this
issue in more detail at the end of the section, but for now we simply
assume a proof \[\begin{array}{ll}
&\!\! \mathsf{Equal^\wedge ArrKT}\\
\quad\quad  : & \!\!\mathsf{Equal^{\wedge} \, A\, (B
  \to C)\, K_\top\, (Arr^{\wedge} \, B\, C\, K_\top \, K_\top)\, e}
\end{array}\]
and define
\[\begin{array}{ll}
 & \!\!\mathsf{LTerm^\wedge KT\,A\, (abs\, B \,C \, e \,s \,T \,
  t')}\\
\quad\quad= & \!\!\mathsf{(K_\top , K_\top , Equal^\wedge ArrKT ,}\\
 & \hspace*{0.3in} \mathsf{LType^\wedge KT\, B\, T , LTerm^\wedge KT\, C\, t') }
\end{array}\]
\item If $\mathsf{t_1 : LTerm\,(B \to A)}$ and $\mathsf{t_2 :
  LTerm\,B}$ then, by the same reasoning as in the previous case, we
  need to define
  \[\begin{array}{ll}
 & \!\! \mathsf{LTerm^\wedge KT\,A\, (app\, B \, t_1\,t_2)}\\
\quad\quad : &\!\! \mathsf{LTerm^{\wedge}\, (B \to A)\, (Arr^{\wedge} \, B\, A\,
  K_\top \, K_\top)\, t_1\; \times}\\
 & \hspace*{1.5in} \mathsf{LTerm^{\wedge}\, B\, K_\top\, t_2}
  \end{array}\]
  We define the second component of the pair to be
  $\mathsf{LTerm^\wedge KT\,B\,t_2}$. We define the first
  component from a proof of $\mathsf{LTerm^{\wedge}\, (B \to A)\,
    K_\top\, t_1}$ and the function\looseness=-1
  \[\begin{array}{ll}
  & \!\! \mathsf{LTerm^\wedge EqualMap}\\
  \quad\quad : &
  \!\!\mathsf{\forall\, (A : Set)\, (Q\,Q' : A \to
    Set) \to}\\
  & \;\;\mathsf{Equal^\wedge\,A\,A\,Q\,Q'\,refl \to}\\
  &  \;\;\;\;\mathsf{PredMap\,(LTerm\,A)\,
    (LTerm^{\wedge}\,A\,Q)\,(LTerm^{\wedge}\,A\,Q')}
  \end{array}\] that takes two
  (extensionally) equal predicates with the same carrier and produces
  a morphism of predicates between their liftings. We define
  $\mathsf{LTerm^\wedge EqualMap}$ straightforwardly by
  pattern matching on the first two arguments to $\mathsf{PredMap}$ in
  its return type, using transitivity and symmetry of the type
  constructor $\mathsf{Equal}$, together with the two analogously defined
  functions $\mathsf{LType^\wedge EqualMap}$ and $\mathsf{Arr^\wedge
    EqualMap}$ in the cases when the first argument to
  $\mathsf{PredMap}$ is constructed using $\mathsf{var}$ and
  $\mathsf{app}$, respectively. If $\mathsf{L_{K_\top} :
    LTerm^{\wedge}\, (B \to A) \, K_\top\, t_1}$ is the proof
  $\mathsf{L_{K_\top} = LTerm^\wedge KT\, (B \to A)\, t_1}$ and
  $\mathsf{ LTerm^\wedge Arr}$ $\mathsf{: LTerm^{\wedge}\, (B \to A)\,
    (Arr^{\wedge} \, B\, A\, K_\top \, K_\top)\, t_1}$ is the proof
  \[\begin{array}{l}
   \quad\quad  \mathsf{LTerm^\wedge Arr = LTerm^\wedge
    EqualMap\,K_\top\,(Arr^{\wedge}\,B\,A\,K_\top\,K_\top)}\\
  \hspace*{1.5in} \mathsf{Equal^\wedge ArrKT\, t_1\, L_{K_\top}}
  \end{array}\]
  then we define
  \[\begin{array}{ll}
    & \mathsf{LTerm^\wedge KT\,A\, (app\, B\, t_1\, t_2)} \\
  = & \mathsf{(K_\top ,  LTerm^\wedge Arr , LTerm^\wedge KT\,B\,t_2)}
    \end{array}\]
\item If $\mathsf{e : Equal\,A\,(List\,B)}$ and $\mathsf{ts :
  List\,(LTerm\, B)}$ then, as above, we need to define
  \[\begin{array}{ll}
  & \!\!\mathsf{LTerm^\wedge KT\,A\, (list\, B \,e\, ts)}\\
\quad\quad : & \!\!\mathsf{Equal^{\wedge} \, A\, (List\,B)\, K_\top\,
  (List^{\wedge} \, B\, K_\top) \, e\, \times}\\
 & \quad\mathsf{List^{\wedge}\, (LTerm\,B)
  \, (LTerm^{\wedge} \, B\, K_\top) \, ts }
  \end{array}\]
  As in that case we assume a proof
  \[\;\;\quad\mathsf{Equal^\wedge ListKT : Equal^{\wedge} \, A\, (List\,B)\, K_\top\,
  (List^{\wedge} \, B\, K_\top) \, e}\] for the first component.  We
  can define the second component using $\mathsf{liftListMap}$ from
  Section~\ref{sec:ind-lam} to map a morphism $\mathsf{PredMap\,
    (LTerm\,B)\, (K_\top)\, (LTerm^{\wedge}\,B\,K_\top)}$ of
  predicates to a morphism $\mathsf{PredMap\,(List\,(LTerm\,B))}$
  $\mathsf{(List^{\wedge}\,(LTerm\,B)\,K_\top) \,
    (List^{\wedge}\,(LTerm\,B)\, (LTerm^{\wedge}\,B\,K_\top))}$ of
  lifted predicates.  Taking
  \[\mathsf{m_{K_\top} : PredMap\,
    (LTerm\,B) \, (K_\top)\, (LTerm^{\wedge}\,B\,K_\top)}\] to be the
  proof \[\mathsf{m_{K_\top} \, t'\, tt\, = LTerm^\wedge KT\, B\, t'}\]
  where $\mathsf{t' : LTerm\,B}$ and $\mathsf{tt}$ is the single
  element of $\mathsf{K_\top\, t'}$, and taking
  \[\mathsf{L_{List^\wedge LTerm^\wedge KT} : List^{\wedge}\,
    (LTerm\,B) \, (LTerm^{\wedge} \, B\, K_\top) \, ts}\] to be the
  proof
\[\begin{array}{ll}
& \!\!\mathsf{ L_{List^\wedge LTerm^\wedge KT}}\\
\quad\quad= & 
 \!\!\mathsf{liftListMap \, (LTerm\,B) \, K_\top \, (LTerm^{\wedge}\,
   B\, K_\top) \, m_{K_\top} \, ts}\\
 & \!\!\quad \mathsf{(List^\wedge KT\, (LTerm\, B)\,
   ts) }
\end{array}\]
we define

\pagebreak

\vspace*{-0.35in}

\[\begin{array}{ll}
 & \!\!\mathsf{LTerm^\wedge KT\,A\, (list\, B\, e\,
  ts)}\\
\quad\quad = & \!\!\mathsf{(K_\top , \, Equal^\wedge ListKT , \, L_{List^\wedge
    LTerm^\wedge KT} ) }
\end{array}\]
\end{itemize}

The above techniques can be used to define a function
$\mathsf{G^\wedge KT : \forall\, (A : Set)\, (x : G\,A) \to
  G^{\wedge}\, A\, K_\top\, x}$ for any GADT $\mathsf{G}$ as defined
in Section~\ref{sec:framework}.  To provide a proof of
$\mathsf{G^{\wedge}\, A\, K_\top \, x}$ for every term $\mathsf{x :
  G\, A}$, we need to know that, if $\mathsf{G}$ has a constructor
$\mathsf{c : H \, (\ol{F\,G\, \ol{B}}) \to G\, (\ol{K\,\ol{B}})}$,
then $\mathsf{H}$ cannot construct a GADT so the generalization
$\mathsf{H^\wedge Map}$ of $\mathsf{listLiftMap}$ in the final bullet
point above is guaranteed to exist. We also need to know that the
lifting of $\mathsf{K_\top}$ to types constructed by any nested type
constructor $\mathsf{F}$ is extensionally equal to $\mathsf{K_\top}$
on the types it constructs. For example, we might need a proof that
$\mathsf{Pair^{\wedge}\,A\,B\,K_\top\,K_\top}$ is equal to
$\mathsf{K_\top}$ on $\mathsf{A \times B}$.  Given a pair $\mathsf{(a
  , b) : A \times B}$, we have that
\[\mathsf{Pair^{\wedge}\,A\,B\,K_\top\,K_\top (a, b) = K_\top \, a
  \times K_\top\, b = \top \times \top}\] whereas $\mathsf{K_\top\,
  (a, b) = \top}$. While these types are not equal, they are clearly
isomorphic. Similar isomorphisms between
$\mathsf{F^{\wedge}\,A\,K_\top}$ and $\mathsf{K_\top}$ hold for all
other nested type constructors $\mathsf{F}$ as well. These
isomorphisms can either be proved on an as-needed basis or, since
$\mathsf{F^\wedge\,A\,K_\top = K_\top}$ is the unary analogue of the
Identity Extension Lemma, be obtained at the meta-level as a
consequence of unary parametricity. At the object level, our Agda
code
%~\cite{web-page}
simply postulates each isomorphism needed since
an Agda implementation of full parametricity for some relevant
calculus is beyond the scope of the present paper.

\section{Conclusion}\label{sec:conclusion}

This paper extends (deep) induction to GADTs that are not truly nested
GADTs. It also shows that truly nested GADTs do not obviously admit
(deep) induction rules. Our development is implemented in Agda, as is
our case study from Section~\ref{sec:app}. Our development opens the
way to incorporating automatic generation of (deep) induction rules
for them into proof assistants.

\subsection*{Acknowledgments}

This work was supported by National Science Foundation award 1906388.

\bibliographystyle{ACM-Reference-Format}
\bibliography{bibfile}


\begin{thebibliography}{28}

%%% the Web address.

\ifx \showCODEN    \undefined \def \showCODEN     #1{\unskip}     \fi
\ifx \showDOI      \undefined \def \showDOI       #1{#1}\fi
\ifx \showISBNx    \undefined \def \showISBNx     #1{\unskip}     \fi
\ifx \showISBNxiii \undefined \def \showISBNxiii  #1{\unskip}     \fi
\ifx \showISSN     \undefined \def \showISSN      #1{\unskip}     \fi
\ifx \showLCCN     \undefined \def \showLCCN      #1{\unskip}     \fi
\ifx \shownote     \undefined \def \shownote      #1{#1}          \fi
\ifx \showarticletitle \undefined \def \showarticletitle #1{#1}   \fi
\ifx \showURL      \undefined \def \showURL       {\relax}        \fi
% invisible to TeX
\providecommand\bibfield[2]{#2}
\providecommand\bibinfo[2]{#2}
\providecommand\natexlab[1]{#1}
\providecommand\showeprint[2][]{arXiv:#2}

\bibitem[\protect\citeauthoryear{Atkey}{Atkey}{2012}]%
        {atk12}
\bibfield{author}{\bibinfo{person}{R. Atkey}.} \bibinfo{year}{2012}\natexlab{}.
\newblock \showarticletitle{Relational parametricity for higher kinds}. In
  \bibinfo{booktitle}{\emph{Computer Science Logic}}. \bibinfo{pages}{46--61}.
\newblock


\bibitem[\protect\citeauthoryear{Bainbridge, Freyd, Scedrov, and
  Scott}{Bainbridge et~al\mbox{.}}{1990}]%
        {bfss90}
\bibfield{author}{\bibinfo{person}{E.~S. Bainbridge}, \bibinfo{person}{P.
  Freyd}, \bibinfo{person}{A. Scedrov}, {and} \bibinfo{person}{P.~J. Scott}.}
  \bibinfo{year}{1990}\natexlab{}.
\newblock \showarticletitle{Functorial polymorphism}.
\newblock \bibinfo{journal}{\emph{Theoretical Computer Science}}
  \bibinfo{volume}{70(1)} (\bibinfo{year}{1990}), \bibinfo{pages}{35--64}.
\newblock
\urldef\tempurl%
\url{https://doi.org/10.1016/0304-3975(90)90151-7}
\showDOI{\tempurl}


\bibitem[\protect\citeauthoryear{Bird and Meertens}{Bird and Meertens}{1998}]%
        {bm98}
\bibfield{author}{\bibinfo{person}{R. Bird} {and} \bibinfo{person}{L.
  Meertens}.} \bibinfo{year}{1998}\natexlab{}.
\newblock \showarticletitle{Nested datatypes}. In
  \bibinfo{booktitle}{\emph{Mathematics of Program Construction}}.
  \bibinfo{pages}{52--67}.
\newblock
\urldef\tempurl%
\url{https://doi.org/10.1007/BFb0054285}
\showDOI{\tempurl}


\bibitem[\protect\citeauthoryear{Cheney and Hinze}{Cheney and Hinze}{2003}]%
        {ch03}
\bibfield{author}{\bibinfo{person}{J. Cheney} {and} \bibinfo{person}{R.
  Hinze}.} \bibinfo{year}{2003}\natexlab{}.
\newblock \bibinfo{title}{First-class phantom types}.  (\bibinfo{year}{2003}).
\newblock
\newblock
\shownote{CUCIS TR2003-1901, Cornell University}.


\bibitem[\protect\citeauthoryear{Dybjer}{Dybjer}{1994}]%
        {dyb94}
\bibfield{author}{\bibinfo{person}{P. Dybjer}.}
  \bibinfo{year}{1994}\natexlab{}.
\newblock \showarticletitle{Inductive families}.
\newblock \bibinfo{journal}{\emph{Formal Aspects of Computing}}
  \bibinfo{volume}{6(4)} (\bibinfo{year}{1994}), \bibinfo{pages}{440--465}.
\newblock
\urldef\tempurl%
\url{https://doi.org/10.1007/BF01211308}
\showDOI{\tempurl}


\bibitem[\protect\citeauthoryear{Fu and Selinger}{Fu and Selinger}{2018}]%
        {fs18}
\bibfield{author}{\bibinfo{person}{P. Fu} {and} \bibinfo{person}{P. Selinger}.}
  \bibinfo{year}{2018}\natexlab{}.
\newblock \bibinfo{title}{Dependently typed folds for nested data types}.
  (\bibinfo{year}{2018}).
\newblock
\urldef\tempurl%
\url{https://arxiv.org/abs/1806.05230}
\showURL{%
\tempurl}


\bibitem[\protect\citeauthoryear{Ghani, Johann, Forsberg, Orsanigo, and
  Revell}{Ghani et~al\mbox{.}}{2015}]%
        {gjfor15}
\bibfield{author}{\bibinfo{person}{N. Ghani}, \bibinfo{person}{P. Johann},
  \bibinfo{person}{F.~Nordvall Forsberg}, \bibinfo{person}{F. Orsanigo}, {and}
  \bibinfo{person}{T. Revell}.} \bibinfo{year}{2015}\natexlab{}.
\newblock \showarticletitle{Bifibrational functorial semantics for parametric
  polymorphism}. In \bibinfo{booktitle}{\emph{Mathematical Foundations of
  Program Semantics}}. \bibinfo{pages}{165--181}.
\newblock
\urldef\tempurl%
\url{https://doi.org/10.1016/j.entcs.2015.12.011}
\showDOI{\tempurl}


\bibitem[\protect\citeauthoryear{Hinze}{Hinze}{2003}]%
        {hin03}
\bibfield{author}{\bibinfo{person}{R. Hinze}.} \bibinfo{year}{2003}\natexlab{}.
\newblock \showarticletitle{Fun with phantom types}. In
  \bibinfo{booktitle}{\emph{The Fun of Programming}}.
  \bibinfo{pages}{245--262}.
\newblock


\bibitem[\protect\citeauthoryear{Johann and Ghani}{Johann and Ghani}{2008}]%
        {jg08}
\bibfield{author}{\bibinfo{person}{P. Johann} {and} \bibinfo{person}{N.
  Ghani}.} \bibinfo{year}{2008}\natexlab{}.
\newblock \showarticletitle{Foundations for Structured Programming with GADTs}.
  In \bibinfo{booktitle}{\emph{Proceedings, Principles of Programming
  Languages}}. \bibinfo{pages}{297--308}.
\newblock
\urldef\tempurl%
\url{https://doi.org/10.1145/1328438.1328475}
\showDOI{\tempurl}


\bibitem[\protect\citeauthoryear{Johann, Ghiorzi, and Jeffries}{Johann
  et~al\mbox{.}}{2021a}]%
        {jg21}
\bibfield{author}{\bibinfo{person}{P. Johann}, \bibinfo{person}{E. Ghiorzi},
  {and} \bibinfo{person}{D. Jeffries}.} \bibinfo{year}{2021}\natexlab{a}.
\newblock \showarticletitle{GADTs, functoriality, parametricity: Pick two}. In
  \bibinfo{booktitle}{\emph{Logical and Semantic Frameworks with
  Applications}}.
\newblock


\bibitem[\protect\citeauthoryear{Johann, Ghiorzi, and Jeffries}{Johann
  et~al\mbox{.}}{2021b}]%
        {jgj21f}
\bibfield{author}{\bibinfo{person}{P. Johann}, \bibinfo{person}{E. Ghiorzi},
  {and} \bibinfo{person}{D. Jeffries}.} \bibinfo{year}{2021}\natexlab{b}.
\newblock \showarticletitle{Parametricity for primitive nested types}. In
  \bibinfo{booktitle}{\emph{Foundations of Software Science and Computation
  Structures}}. \bibinfo{pages}{324--343}.
\newblock
\urldef\tempurl%
\url{https://doi.org/10.1007/978-3-030-71995-1\_17}
\showDOI{\tempurl}


\bibitem[\protect\citeauthoryear{Johann and Polonsky}{Johann and
  Polonsky}{2019}]%
        {jp19}
\bibfield{author}{\bibinfo{person}{P. Johann} {and} \bibinfo{person}{A.
  Polonsky}.} \bibinfo{year}{2019}\natexlab{}.
\newblock \showarticletitle{Higher-kinded data types: Syntax and semantics}. In
  \bibinfo{booktitle}{\emph{Logic in Computer Science}}.
  \bibinfo{pages}{1--13}.
\newblock
\urldef\tempurl%
\url{https://doi.org/10.1109/LICS.2019.8785657}
\showDOI{\tempurl}


\bibitem[\protect\citeauthoryear{Johann and Polonsky}{Johann and
  Polonsky}{2020}]%
        {jp20}
\bibfield{author}{\bibinfo{person}{P. Johann} {and} \bibinfo{person}{A.
  Polonsky}.} \bibinfo{year}{2020}\natexlab{}.
\newblock \showarticletitle{Deep induction: Induction rules for (truly) nested
  types}. In \bibinfo{booktitle}{\emph{Foundations of Software Science and
  Computation Structures}}. \bibinfo{pages}{339--358}.
\newblock
\urldef\tempurl%
\url{https://doi.org/10.1007/978-3-030-45231-5_18}
\showDOI{\tempurl}


\bibitem[\protect\citeauthoryear{Jones, Vytiniotis, Weirich, and
  Washburn}{Jones et~al\mbox{.}}{2006}]%
        {pvww06}
\bibfield{author}{\bibinfo{person}{S.~Peyton Jones}, \bibinfo{person}{D.
  Vytiniotis}, \bibinfo{person}{S. Weirich}, {and} \bibinfo{person}{G.
  Washburn}.} \bibinfo{year}{2006}\natexlab{}.
\newblock \showarticletitle{Simple unification-based type inference for GADTs}.
  In \bibinfo{booktitle}{\emph{International Conference on Functional
  Programming}}. \bibinfo{pages}{50--61}.
\newblock
\urldef\tempurl%
\url{https://doi.org/10.1145/1160074.1159811}
\showDOI{\tempurl}


\bibitem[\protect\citeauthoryear{Lane}{Lane}{1971}]%
        {mac71}
\bibfield{author}{\bibinfo{person}{S.~Mac Lane}.}
  \bibinfo{year}{1971}\natexlab{}.
\newblock \bibinfo{booktitle}{\emph{Categories for the Working Mathematician}}.
\newblock \bibinfo{publisher}{Springer}.
\newblock


\bibitem[\protect\citeauthoryear{McBride}{McBride}{1999}]%
        {mcb99}
\bibfield{author}{\bibinfo{person}{C. McBride}.}
  \bibinfo{year}{1999}\natexlab{}.
\newblock \bibinfo{title}{Dependently Typed Programs and their Proofs}.
  (\bibinfo{year}{1999}).
\newblock
\newblock
\shownote{PhD thesis, University of Edinburgh}.


\bibitem[\protect\citeauthoryear{Minsky}{Minsky}{2015}]%
        {min15}
\bibfield{author}{\bibinfo{person}{Y. Minsky}.}
  \bibinfo{year}{2015}\natexlab{}.
\newblock \bibinfo{title}{Why GADTs Matter for Performance}.
  (\bibinfo{year}{2015}).
\newblock
\urldef\tempurl%
\url{https://blogs.janestreet.com/why-gadts-matter-for-performance}
\showURL{%
\tempurl}


\bibitem[\protect\citeauthoryear{Pasalic and Linger}{Pasalic and
  Linger}{2004}]%
        {pl04}
\bibfield{author}{\bibinfo{person}{E. Pasalic} {and} \bibinfo{person}{N.
  Linger}.} \bibinfo{year}{2004}\natexlab{}.
\newblock \showarticletitle{Meta-programming with typed object-language
  representations}. In \bibinfo{booktitle}{\emph{Generic Programming and
  Component Engineering}}. \bibinfo{pages}{136--167}.
\newblock
\urldef\tempurl%
\url{https://doi.org/10.1007/978-3-540-30175-2\_8}
\showDOI{\tempurl}


\bibitem[\protect\citeauthoryear{Pottier and R{\'e}gis-Gianas}{Pottier and
  R{\'e}gis-Gianas}{2006}]%
        {pr06}
\bibfield{author}{\bibinfo{person}{F. Pottier} {and} \bibinfo{person}{Y.
  R{\'e}gis-Gianas}.} \bibinfo{year}{2006}\natexlab{}.
\newblock \showarticletitle{Stratified type inference for generalized algebraic
  data types}. In \bibinfo{booktitle}{\emph{Principles of Programming
  Languages}}. \bibinfo{pages}{232--244}.
\newblock
\urldef\tempurl%
\url{https://doi.org/10.1145/1111320.1111058}
\showDOI{\tempurl}


\bibitem[\protect\citeauthoryear{Roundy}{Roundy}{2006}]%
        {rou06}
\bibfield{author}{\bibinfo{person}{D. Roundy}.}
  \bibinfo{year}{2006}\natexlab{}.
\newblock \bibinfo{title}{Implementing the darcs Patch Formalism ...and
  Verifying It}.  (\bibinfo{year}{2006}).
\newblock
\urldef\tempurl%
\url{https://physics.oregonstate.edu/~roundyd/talks/fosdem}
\showURL{%
\tempurl}


\bibitem[\protect\citeauthoryear{Schrijvers, Jones, Sulzmann, and
  Vytiniotis}{Schrijvers et~al\mbox{.}}{2009}]%
        {sjsv09}
\bibfield{author}{\bibinfo{person}{T. Schrijvers},
  \bibinfo{person}{S.~L.~Peyton Jones}, \bibinfo{person}{M. Sulzmann}, {and}
  \bibinfo{person}{D. Vytiniotis}.} \bibinfo{year}{2009}\natexlab{}.
\newblock \showarticletitle{Complete and decidable type inference for GADTs}.
  In \bibinfo{booktitle}{\emph{International Conference on Functional
  Programming}}. \bibinfo{pages}{341--352}.
\newblock
\urldef\tempurl%
\url{https://doi.org/10.1145/1631687.1596599}
\showDOI{\tempurl}


\bibitem[\protect\citeauthoryear{Sheard and Pasalic}{Sheard and
  Pasalic}{2004}]%
        {sp04}
\bibfield{author}{\bibinfo{person}{T. Sheard} {and} \bibinfo{person}{E.
  Pasalic}.} \bibinfo{year}{2004}\natexlab{}.
\newblock \showarticletitle{Meta-programming with built-in type equality}. In
  \bibinfo{booktitle}{\emph{Workshop on Logical Frameworks and
  Meta-languages}}. \bibinfo{pages}{106--124}.
\newblock


\bibitem[\protect\citeauthoryear{Tassi}{Tassi}{2019}]%
        {tas19}
\bibfield{author}{\bibinfo{person}{E. Tassi}.} \bibinfo{year}{2019}\natexlab{}.
\newblock \showarticletitle{Deriving proved equality tests in Coq-elpi:
  Stronger induction principles for containers in Coq}. In
  \bibinfo{booktitle}{\emph{Interactive Theorem Proving}}.
  \bibinfo{pages}{1--18}.
\newblock
\urldef\tempurl%
\url{https://doi.org/10.4230/LIPIcs.CVIT.2016.23}
\showDOI{\tempurl}


\bibitem[\protect\citeauthoryear{Team}{Team}{2020}]%
        {coq20}
\bibfield{author}{\bibinfo{person}{The Coq~Development Team}.}
  \bibinfo{year}{2020}\natexlab{}.
\newblock \bibinfo{title}{The Coq Proof Assistant, version 8.11.0}.
  (\bibinfo{year}{2020}).
\newblock
\urldef\tempurl%
\url{https://doi.org/10.5281/zenodo.3744225}
\showDOI{\tempurl}


\bibitem[\protect\citeauthoryear{Ullrich}{Ullrich}{2020}]%
        {ull20}
\bibfield{author}{\bibinfo{person}{M. Ullrich}.}
  \bibinfo{year}{2020}\natexlab{}.
\newblock \bibinfo{title}{Generating Induction Principles for Nested Induction
  Types in MetaCoq}.  (\bibinfo{year}{2020}).
\newblock
\newblock
\shownote{PhD thesis, Saarland University}.


\bibitem[\protect\citeauthoryear{Vytiniotis and Weirich}{Vytiniotis and
  Weirich}{2010}]%
        {vw10}
\bibfield{author}{\bibinfo{person}{D. Vytiniotis} {and} \bibinfo{person}{S.
  Weirich}.} \bibinfo{year}{2010}\natexlab{}.
\newblock \showarticletitle{Parametricity, type equality, and higher-order
  polymorphism}.
\newblock \bibinfo{journal}{\emph{Journal of Functional Programming}}
  \bibinfo{volume}{20(2)} (\bibinfo{year}{2010}), \bibinfo{pages}{175--210}.
\newblock
\urldef\tempurl%
\url{https://doi.org/10.1017/S0956796810000079}
\showDOI{\tempurl}


\bibitem[\protect\citeauthoryear{Xi, Chen, and Chen}{Xi et~al\mbox{.}}{2003}]%
        {xcc03}
\bibfield{author}{\bibinfo{person}{H. Xi}, \bibinfo{person}{C. Chen}, {and}
  \bibinfo{person}{G. Chen}.} \bibinfo{year}{2003}\natexlab{}.
\newblock \showarticletitle{Guarded recursive datatype constructors}. In
  \bibinfo{booktitle}{\emph{Principles of Programming Languages}}.
  \bibinfo{pages}{224--235}.
\newblock
\urldef\tempurl%
\url{https://doi.org/10.1145/604131.604150}
\showDOI{\tempurl}


\bibitem[\protect\citeauthoryear{Zilberstein}{Zilberstein}{2015}]%
        {cis194}
\bibfield{author}{\bibinfo{person}{N. Zilberstein}.}
  \bibinfo{year}{2015}\natexlab{}.
\newblock \bibinfo{title}{CIS194 homepage}.  (\bibinfo{year}{2015}).
\newblock
\urldef\tempurl%
\url{https://www.seas.upenn.edu/~cis194/spring15/lectures/11-stlc.html}
\showURL{%
\tempurl}


\end{thebibliography}

\end{document}